\documentclass[prepint, 5p, twocolumn]{elsarticle}

\usepackage{amsmath,amssymb,amsfonts}
\usepackage{mathtools}
\usepackage{algorithmic}
\usepackage{graphicx}
\usepackage{xcolor}
\usepackage{enumerate}
\usepackage{adjustbox}
\usepackage{booktabs}
\usepackage{adjustbox}
\usepackage{textcomp}
\usepackage{multirow}
\usepackage{cuted}
\usepackage{xstring}
\usepackage[hidelinks]{hyperref}
\newtheorem{Def}{Definition}
\newtheorem{Theo}{Theorem}
\newtheorem{Cor}{Corollary}
\newtheorem{Ex}{Example}

\newtheorem{Assum}{Assumption}
\newtheorem{prob}{Problem}
\newproof{pf}{Proof}
\numberwithin{Def}{section}
\numberwithin{Theo}{section} 
\numberwithin{Cor}{section}
\numberwithin{Ex}{section} 
\numberwithin{Prop}{section}
\numberwithin{Assum}{section}

\newcommand{\sch}[4]{\ensuremath{#1 - #2#3^{-1}#4}}

\providecommand{\com[2]}{\begin{tt}[#1: #2]\end{tt}}
\definecolor{red}{rgb}{1,0,0}

\definecolor{blu}{rgb}{0,0,1}

\definecolor{gre}{rgb}{0,0.7,0.3}

\definecolor{cyan2}{cmyk}{1, 0, 0, 0.3}

\begin{document}

\begin{frontmatter}
	
	\title{Optimal excitation and measurement pattern for cascade networks \tnoteref{footnoteinfo}} 
	
	\tnotetext[footnoteinfo]{This study was financed in part by the Coordenação de Aperfeiçoamento de Pessoal de Nível Superior - Brasil (CAPES) - Finance Code 001. This paper was not presented at any IFAC Meeting. 
	Corresponding author E.~Mapurunga.}
	 
	\author[First]{Eduardo Mapurunga} 
	\author[First]{Alexandre Sanfelice Bazanella} 
	\address[First]{Data-Driven Control Group \\ Department of Automation and Energy, Universidade Federal do Rio Grande do Sul (DELAE/UFRGS), Porto Alegre-RS, Brazil.
		\\email: \{eduardo.mapurunga, bazanella\}@ufrgs.br}

	
	\begin{abstract}                
			This work deals with accuracy analysis of dynamical systems interconnected in a cascade structure. 
			For a cascade network there are a number of experimental settings for which the dynamic systems within the network can be identified. 
			We study the problem of choosing which excitation and measurement pattern delivers the most accurate parameter estimates for the whole network.  
			The optimal experiment is based on the accuracy assessed through the asymptotic covariance matrix of the prediction error method, 
			while the cost criterion is the number of excitations and measurements.  
			We develop theoretical results under the assumptions that all dynamic systems are equal and with equal signal-to-noise ratio throughout the network.   
			We show that there are experimental settings which result in equal overall precision and that there is an excitation and measurement pattern that yields more accurate results than others.
		 	From these results a guideline based on the topology of the network emerges for the choice of the experimental setting. 
			We provide numerical results which attest that the principles behind this guideline are also valid for more general situations. 
	\end{abstract}
	
	\begin{keyword}
		Dynamic Networks, Network Identification, Variance Analysis.
	\end{keyword}
	
\end{frontmatter}

\section{Introduction}

Interconnected systems, or dynamic networks, are becoming a common framework for problems in science and engineering. 
Dynamic networks find a wide range of applications:
from epidemic control \citep{gracy_analysis_2021} to energy storage systems \citep{Han_nextgeneration_2020}.
A network can be represented by node signals, usually variables of interest, and modules which define the dynamic relationship among the nodes of the network. 
Obtaining reliable models for these networks can lead to advances in many areas, such as biomedical and social sciences.  
Identification methods that allow to obtain such models are of paramount importance and they are currently an active field of research. 

Identification in dynamic networks can be roughly divided into three categories: identification of a single module embedded in a network, topology detection, and identification of the whole network.
Several methods have been proposed for identification of a single module from a dynamic network in \cite{VanDenHof2013ComplexPem,dankers_errors--variables_2015, Dankers2016PredInputSel,gevers_practical_2018, ramaswamy_local_2020, ramaswamy_learning_2021}.  
Topology detection was addressed in \cite{YUAN20111230, dimovska_control_2021, jin_full_2021, van_waarde_topology_2021}.
Identification of the whole network can be addressed by the prediction error method \citep{weerts_prediction_2018}, while some other methods have been proposed in \cite{weerts_sequential_2018}. 
The focus of this work is on identifying all modules of a dynamic network with cascade topology. 

For identification methods to provide consistent estimates of some modules it is necessary
that they can be uniquely determined from the network data - that is, that the modules
are {\em identifiable}. 
The question of whether some modules, or the full network, are identifiable has been explored in \cite{Gevers2015ExpdsgIssues, weerts_identifiability_2015, Gevers2017IdentifiDyNet, Bazanella2017WhichNodes} and for the rank-reduced noise case in \cite{weerts_identifiability_2018, gevers_identifiability_2019}.
Identifiability of the modules is determined primarily by the location of excitations and measurements in the network.
A recent research direction has been to characterize how to allocate input and output signals in a dynamic network to uniquely recover its modules \citep{Hendrickx2018PartialNodes, cheng_allocation_2021, van_waarde_necessary_2019, bazanella_network_2019, legat_local_2020, weerts_abstractions_2020, mapurunga_identifiability_2021}.
These results open a number of possibilities for excitation and measurement patterns from which a user must choose.  
Therefore, tools are necessary to guide the user's choice on the best experimental setting according to her objectives.

Once identifiability of the network is determined, the accuracy of the parameter estimates obtained by the identification method becomes the next natural concern. 
Accuracy of the parameter estimates depends on many factors: the parametrization of the modules, the signal-to-noise ratio (SNR) applied in the network, and on the location of the inputs and measurements taken from the network.
In this work we explore the problem of how to choose the excitation and measurement pattern (EMP) in order to obtain the most accurate parameter estimates for cascade networks using the minimum number of excitations and measurements. 
Cascade systems find good applications in engineering and industrial processes \citep{Wahlberg2007}. 
Results obtained for cascade networks can be applied to more general topologies, such as trees and directed acyclic graphs, for which a cascade structure are a part. 
For cascade networks, we explore how the trade-off among the excitations and measurements at some nodes influence the accuracy of the EMPs. 
Moreover, the contribution of some modules to the precision of the parameter estimates considering the experimental setting is investigated.

Deciding which nodes to excite and which nodes to measure represent a structural problem in the experiment design. 
In other words, the accuracy of the module estimates will also depend on the distribution of the exciting signals and measurements from the network rather than only their associated energy. 
Even if the user can design a constrained input to improve the quality of the obtained model,  
the accuracy of the model may be severely compromised only due to the distribution of external excitations and measurements in the network. 
Therefore, the structure of the EMP of a network has an impact on how good the performance of the identification method can be.

The problem of how to select the location of inputs and outputs so to obtain optimal parameter estimates has been almost completely unexplored in the literature. 
It has already been investigated in \cite{mapurunga2021sysid} for cascade and cycle networks, but the authors have specialized the analysis for state space networks.
Interesting patterns were pointed out; specifically, for cascade networks with $n$ modules, it has been found that it is best to excite the first $n/2$ nodes and measure the remaining.
In \cite{Wahlberg2009CascAuto} accuracy aspects of modules in a cascade network were also investigated, 
but their results were specific for one particular EMP.

In this work we aim to determine guidelines to choose the EMP that provides the most accurate identification. 
We first isolate the effect of the EMP on the accuracy by considering the case in which all modules and all  signal-to-noise ratios in the network are the same.
For this case, analytical results are obtained that indicate the best EMPs with minimal cardinality, besides revealing other relevant properties. 
The results for cascade networks found in \cite{mapurunga2021sysid} are extended to modules represented by rational transfer functions; 
the results of \cite{Wahlberg2009CascAuto} are also extended in several directions.  
Then we consider the effect on the accuracy of the structure and size of the modules, and of the signal to noise ratios,  and how this influences the choice of the EMP. 
This is done mostly by exhaustive numerical simulations and indicate that, to a large extent, the guidelines derived from the analytical results previously obtained can be generalized to this situation where all modules are different and when there is unequal SNRs applied at each node.

The paper is organized as follows. 
In Section \ref{sec:Problem} we present the network setup considered in this work and we formally state the problem that will be tackled. 
We start our analysis with cascade networks composed of three nodes in Section \ref{sec:3nodes}.
In Section \ref{sec:4nodes} we extend our results to four nodes cascade networks. 
These results are further explored in cascade networks with an arbitrary number of nodes in section \ref{sec:Nnodes}. 
Numerical examples are reported in Section \ref{sec:Numerical} and the conclusions are presented in Section \ref{sec:Conclusion}. 

\section{Problem statement}
\label{sec:Problem}

Cascade systems are dynamic networks with a well-defined topology: a branch.
This topology can be represented by a path graph with nodes defined by $\mathcal{W} = \{1, 2, \dots, n\}$ and edges $\mathcal{E} = \{\left(1, 2 \right), \left(2, 3 \right), \dots, \left(n-1, n \right) \}$. 
Each node represents an internal scalar signal of the network, while each edge is represented by a discrete time transfer function, also called module. 
Some nodes may be subjected to known external excitation signals and just a subset of them may be available for direct measurement. 
The dynamics of the cascade network is given by the standard module representation \citep{VanDenHof2013ComplexPem}:
\begin{subequations}
	\begin{align}
		w(t) &= G^0(q)w(t) + Br(t), \label{eq:dynet1}\\
		y(t) &= Cw(t) + e(t), \label{eq:dynet2}
	\end{align}
\end{subequations}
where $w(t) \in \mathbb{R}^n$ represents the internal signals of the cascade network, 
$r(t) \in \mathbb{R}^m $ is the vector of external excitation signals, 
$y(t) \in \mathbb{R}^p$ is the vector of available measurements of the network corrupted by sensor noise $e(t) \in \mathbb{R}^{p}$. 
The transfer matrix $G^0(q)$ is referred to as network matrix and 
for a cascade network it has the form:
\begin{align}
	G^0(q) = \begin{bmatrix}
  0 & 0 \\
	diag(G^0_1(q), G_2^0(q), \dots, G_{n-1}^0(q)) & 0
\end{bmatrix} 
\label{eq:GCasc},
\end{align}
where $q$ is the forward shift operator, i.e. $qr(t) = r(t+1)$, and $diag(\cdot)$ refers to a diagonal matrix with the argument as main diagonal. 
The matrices $B$ and $C$ are selection matrices responsible for indicating where the inputs are applied in the network and which measurements are taken. 
They have in common the property that each row has at most one $1$, while the other entries of the row are filled with zeros.  
Associated with these matrices are the set $\mathcal{B}$ of excited nodes and the set $\mathcal{C}$ of measured nodes. 
We adopt the following assumptions about the modules, input signals and measurement noise throughout the paper:
\begin{enumerate}[(a)]
				\item the transfer functions $G^0_{k}(q)$, for $k = 1, \dots, n-1$ are proper and $T^0(q) \triangleq (I - G^0(q))^{-1}$ is stable; \label{item:GandTstable}
				\item the external signals $\{r_i(t)\}$ are independent zero mean white noise processes with variance $\sigma_i^2$ and uncorrelated with all noise processes $\left\lbrace e_j(t)\right\rbrace$; \label{item:referencesignals}
				\item the corrupting noise sequences $\left\lbrace e_j(t)\right\rbrace $ are independent stationary Gaussian white noise processes with zero mean and variance $\lambda_j$.  \label{item:NoiseGaussian}
\end{enumerate}

We rewrite the network model (\ref{eq:dynet1})-(\ref{eq:dynet2}) into an input-output representation:
\begin{align}
				y(t) = C T^0(q) B r(t) + e(t),\, T^0(q) \triangleq (I - G^0(q))^{-1}
				\label{eq:ToIO}
.\end{align}
The modules associated with the network matrix can be identified from input-output data $\{r(t), y(t)\}$, $t = 1, \dots, N$.
Whether the modules can be uniquely recovered from input-output data will depend on the choice of the matrices $B$ and $C$. 
There may be many combinations of these matrices such that the whole network is identifiable.  
For this purpose, we use the following definition from \cite{mapurunga2021sysid}. 
\begin{Def}
  Let $G(q)$ be a network matrix for which different selection matrices are considered.  
	A pair of selection matrices $B$ and $C$, with its corresponding node sets $\mathcal{B}$ and $\mathcal{C}$, is called an \emph{excitation and measurement pattern} - EMP for short.
	An EMP is said to be valid if it is such that the network (\ref{eq:dynet1})-(\ref{eq:dynet2}) is generically identifiable
	\footnote{A network is generically identifiable if it is identifiable for \textit{almost all} parameters, except for those in a set of measure zero. See \cite{Hendrickx2018PartialNodes} for a thorough treatment.}. 
  Let $\nu = |\mathcal{B}| + |\mathcal{C}|$ \footnote{$|\cdot|$ - Denote cardinality of a set.} be the cardinality of an EMP. 
  A given EMP is said to be \emph{minimal} if it is valid and there is no other valid EMP with smaller cardinality.  
  \label{def:EMP}
\end{Def}

Our interest is to determine which minimal EMP -- one with the minimum number of excitations and measurements combined -- yields the most accurate estimates. 
The set of candidate minimal EMPs is characterized by 
the conditions formally stated in the next Corollary. 

\begin{Cor}[\cite{bazanella_network_2019}]
				In a cascade network (\ref{eq:dynet1})-(\ref{eq:dynet2}) with network matrix (\ref{eq:GCasc}) an EMP is minimal if and only if  
				$1 \in \mathcal{B}$,
				$n \in \mathcal{C}$,
				$\mathcal{B} \cup \mathcal{C} = \mathcal{W}$,  
				$\mathcal{B} \neq \mathcal{C}$.  
				\label{cor:IdentifiabilityConditions}
\end{Cor}

These conditions state that the first node (source) needs to be excited, the last node (sink) needs to be measured, while every other node needs to be either excited or measured. 
Hence, there are a total of $2^{n-2}$ minimal EMPs available for a cascade network with $n$ nodes. 

Once these conditions are fulfilled we need to obtain the parameter estimates through an identification method. 
Identification of the modules is performed with the prediction error method (PEM), which is used to estimate a parametrized model of the cascade network (\ref{eq:ToIO}):
\begin{align}
				y(t, \theta) = C \left( I - G(q, \theta) \right)^{-1} B r(t) + e(t)
				\label{eq:IOpar}
.\end{align}
with each module in $G(q, \theta)$ parametrized independently: $G_k(q, \theta_k)$ for $k = 1, 2, \dots n - 1$.
The one-step ahead predictor of $y(t)$ will be:
\begin{align}
				\hat{y}(t| t-1, \theta) = C \left( I - G(q, \theta) \right)^{-1} B r(t) 
				\label{eq:predictor}
.\end{align}
We assume that there exist a parameter vector $\theta^0$ such that $G(q, \theta^0) \equiv G^0(q)$. 
Our interest is to assess accuracy of the parameter estimates $\hat{\theta}_k$ from input-output data $\{ r(t), y(t)\}$ using PEM.
Recall that the user has freedom to choose the location of the excitation signals and measurements through $B$ and $C$. 
It is well-known that PEM achieves asymptotically the Cramer-Rao lower bound under the Gaussian assumption \citep{ljung1998system}. 
The asymptotic covariance matrix of PEM can be evaluated as: 
\begin{align}
				P = [\mathbb{E} \psi(t, \theta) \Lambda^{-1} \psi^T(t, \theta)]^{-1},	\label{eq:Ppem}
\end{align}
where $\mathbb{E}$ denotes mathematical expectation, $\psi(t, \theta)$ is the gradient of the prediction error ($y(t) - \hat{y}(t|t-1, \theta)$) with respect to the parameter vector $\theta$, and $\Lambda$ is the noise covariance matrix. 
We will be concerned with the choice of the minimal EMP that achieves the most accurate parameter estimates according to a measure of the covariance matrix $P$.
We are now ready to formally state the problem tackled in this paper.

\begin{prob}
				Given a cascade network satisfying the technical conditions (a)-(c), 
				determine which minimal EMP provides the smallest trace of the asymptotic covariance matrix $P$. 
				\label{pro:minEMPacc}
\end{prob}

Here we adopted the trace of $P$ as a criterion, which in the literature of optimal experiment design is known as A-optimally criterion \citep{pukelsheimOptimalDesignExperiments2006}. 
We remark that  similar conclusions can be derived if one considers the determinant of $P$, which is known as D-optimality criterion.

There are various factors that compete against each other to determine which EMP leads to the most accurate estimates: 
the parametrization of the modules, the location of poles and zeros of each transfer function, signal-to-noise ratio at some nodes. 
In order to isolate these factors and provide meaningful insights with respect to the choice of the minimal EMPs, 
the following assumptions are instrumental in the theoretical analysis provided. 
\begin{Assum}
				All transfer functions of the cascade network are \emph{identical}, i.e. $G_k(e^{j \omega}) \equiv G(e^{j \omega})$, for $k = 1, 2, 3, \dots, n-1$. 
				\label{assu:identicalG}
\end{Assum}
\begin{Assum}
				The external excitation signals $\{r_i(t)\}$ have the same variance $\sigma^{2}_i = \sigma^2$ for $i = 1, 2, \dots m$.
				The covariance matrix $\Lambda$ associated with the noise $e(t)$ can be written as $\lambda I_{p}$, where $I_{p}$ is the identify matrix of size $p$. 
				\label{assu:equally}
\end{Assum}

Notice that Assumption \ref{assu:equally} implies that all excitation signals have the same second-order statistical properties, the same is valid for the corruption noise.  
These assumptions may seem restrictive, but they are necessary to isolate particularities of the modules from the structural property of the experimental setting. 
Furthermore, a fair comparison should consider an equally exciting scenario for all EMPs. 
This is precisely what Assumption \ref{assu:equally} does. 
Regarding this assumption, 
we will introduce the following definition that will be useful later on. 

\begin{Def}
				The signal-to-noise ratio from excitation in node $i$ to measurement taken from node $j$, denoted as \emph{SNRji}, is defined as $\sigma_i^2 / \lambda_j$.	
				\label{def:SNRji}
\end{Def}

This definition is related to the ratio of measurement noise $\{e_j(t)\}$ in node $j$ with respect to the input energy from excitation $\{r_i(t)\}$. 
The next definition concerns the excitation and measurement of a particular module. 

\begin{Def}
				A module $G_{i}$ is called a \emph{direct module} if $i \in \mathcal{B}$ and $i+1 \in \mathcal{C}$. 
				\label{def:dirmod}
\end{Def}

In \cite{mapurunga2021sysid} direct modules were pointed out as one of the key factors for determining which minimal EMP provided the most accurate results for cascade and cyclic state space networks.


\section{Choosing between exciting and measuring}
\label{sec:3nodes}

We start our analysis by looking into cascade networks with three nodes. 
A cascade network with three nodes has the following network matrix. 
\begin{align}
				G(q, \theta) = \left[\begin{matrix} 0 & 0 & 0 \\ G_1(q, \theta_1) & 0 & 0 \\ 0 & G_2(q, \theta_2) & 0 \end{matrix} \right]. \label{eq:G3}		
\end{align}
This dynamic network has two modules $G_1(q, \theta_1)$ and $G_2(q, \theta_2)$ to be identified.  
From the identifiability conditions given in Corollary \ref{cor:IdentifiabilityConditions}, we know that the first node needs to be excited and the last node needs to be measured. 
The second node could be either excited or measured, and this defines the two minimal EMPs for this network:
\begin{enumerate}[I.]
				\item $\mathcal{B} = \{1\}, \mathcal{C} = \{2, 3\}$; \label{item:EMP3n2}
				\item $\mathcal{B} = \{1, 2\}, \mathcal{C} = \{3\}$. \label{item:EMP3n1}
\end{enumerate}
We must decide between these two minimal EMPs, or equivalently, choose whether to excite or measure the second node.
In \cite{Wahlberg2009CascAuto} the minimal EMP \ref{item:EMP3n2} was analyzed and some insights were given assuming \emph{identical} transfer functions $(G_1(q) \equiv G_2(q)$), i.e.  
under Assumption \ref{assu:identicalG}. 
Under this premise, the covariance matrix for 
EMP I was derived in \cite{Wahlberg2009CascAuto} as:
\begin{align}
				P_1 = \left[
		\begin{matrix}
			A^{-1} & -A^{-1} \\
			-A^{-1} & A^{-1} + B^{-1}
		\end{matrix}\right], \label{eq:3nodesP_2}
\end{align}
where
\begin{align*}
				A &\triangleq \frac{N}{\lambda_2} \mathbb{E} [G'_1 r_1 \times G_1'^T r_1], \\ 
				B &\triangleq \frac{N}{\lambda_3} \mathbb{E} [G'_2 G_1 r_1 \times G_2'^T G_1 r_1 ]
.\end{align*}
From now on, we drop the arguments $q$ and $t$ for space purposes whenever it seems necessary. 
The prime denotes differentiation with respect to the parameter vectors $\theta_k$, $k = 1, 2$. 
The subscript in $P_1$ is used to refer to minimal EMP \ref{item:EMP3n2}, we will adopt this convention for the rest of the paper. 
Under Assumption \ref{assu:identicalG}, the main conclusions drawn in \cite{Wahlberg2009CascAuto} were 
\begin{itemize}
				\item 	The quality of the estimate \(\hat{\theta}_1\)  is not improved by measurement of $\{y_3(t)\}$.
\item		The covariance of estimate \(\hat{\theta}_2\) is larger than or equal to covariance of \(\hat{\theta}_1\). 
\end{itemize}

The first observation means that one can not improve the quality of the first module estimates by also measuring the sink node. 
The second observation was already pointed out in \cite{mapurunga2021sysid}, where it has been observed that direct modules -- in this case $G_1$, see Definition \ref{def:dirmod} -- are estimated more accurately.  

Before tackling the problem of whether to excite or measure node 2 in this network, let us
provide a dual analysis for minimal EMP II under Assumption \ref{assu:identicalG}. 
The gradient of the minimal EMP II is as follows:
\begin{align}
				\psi_2(t) = \left[
				\begin{matrix}
				G'_1(q, \theta_1) G_2(q, \theta_2) r_1(t) \\
				G'_2(q, \theta_2)  G_1(q, \theta_1) r_1(t) + G_2'(q, \theta_2) r_2(t)
				\end{matrix}
				\right].
				\label{eq:gradiEMP3n1}	
\end{align}
The asymptotic covariance matrix can be found as the inverse of the information matrix:
\begin{align}
				cov\left(\left[\begin{matrix} \hat{\theta}_1 \\ \hat{\theta}_2 \end{matrix}\right]\right) \sim M_2^{-1}, \label{eq:covinfmatrixEMP_3n_1}
\end{align}
where
\begin{align}
	M_2 = \left[
  \begin{matrix}
    F & H \\
    H^T & B + L
  \end{matrix}
	 \right],	\label{eq:M1_3nodes}	
\end{align}

\begin{align*}
				F &\triangleq \frac{N}{\lambda_3} \mathbb{E} [G'_1 G_2 r_1 \times G_1'^T G_2 r_1], \\
				H &\triangleq \frac{N}{\lambda_3} \mathbb{E} [G'_2 G_1 r_1 \times G_1'^T G_2 r_1], \\
				L &\triangleq \frac{N}{\lambda_3} \mathbb{E} [G'_2 r_2 \times G_2'^T r_2]. 
\end{align*}

If the two modules are identical then \(F = H = B\). 
In this case, the asymptotic covariance matrix is:
\begin{align}
		P_2 = \left[ 
\begin{matrix}
	L^{-1} + B^{-1} & -L^{-1} \\
	-L^{-1} &  L^{-1}
\end{matrix}
\right]	\label{eq:3nodesP_1}	
.\end{align}

Hence, under Assumption \ref{assu:identicalG}, we can make the following observations with respect to  
minimal EMP \ref{item:EMP3n1}:

\begin{itemize}
				\item  
					 The covariance of estimate \(\hat{\theta}_1\) is larger than or equal to the covariance of \(\hat{\theta}_2\).
	 				\item  The quality of \(\hat{\theta}_2\) is not improved by excitation $\{r_1\} $, since $cov(\hat{\theta}_2)$ does not depend on $\{r_1\}$ - see (\ref{eq:3nodesP_1}). 
\end{itemize}

These conclusions are dual to those for the EMP I, in this case we are analyzing the effect
of the addition of an excitation source instead of a new measurement.
The measurement $\{y_3\}$ in EMP I is to $cov(\hat{\theta}_1)$ as the excitation signal $\{r_1\}$ in EMP II is to $cov(\hat{\theta}_2)$.  
Furthermore, we observe again that the direct module, in this case $G_2$, is estimated more precisely. 
These conclusions will also be observed for cascade networks with more nodes, by maintaing the structures of these two EMPs.
In EMP I we only excite the source and measure the remaining nodes, and for EMP II we only measure the sink and excite all other nodes. 
In fact, Assumption \ref{assu:identicalG} is not necessary for this phenomenon to happen as we will see in Section \ref{sec:Nnodes}.

The next theorem allows to 
decide whether to measure or excite node two based on the trace of the covariance matrix.  

\begin{Theo}
				Consider a 3-nodes cascade network with network matrix given in (\ref{eq:G3}).  
				Under Assumption \ref{assu:identicalG},  
				EMP II yields a smaller trace of the covariance matrix if and only if SNR32 in EMP II is larger than SNR21 in EMP I: 
				\begin{align}
								\frac{\sigma_2^2}{\lambda_3} > \frac{\sigma_1^2 }{\lambda_2}, \label{eq:snrEMP_1vsEMP_2} 
				\end{align}
				otherwise EMP I is more accurate. 
				Under Assumptions \ref{assu:identicalG} and \ref{assu:equally}, it holds that both EMPs result in the same trace of the covariance matrix.
				\label{theo:EMPsIandIItr3nodes}
\end{Theo}
\begin{pf}
				According to (\ref{eq:3nodesP_2}) and (\ref{eq:3nodesP_1}), the trace of the covariance matrices of the minimal EMPs are:
				\begin{align}
								tr(P_1) &= tr(A^{-1}) + tr(A^{-1} + B^{-1}). \label{eq:3nodesTrP_2}\\
								tr(P_2) &= tr(L^{-1}) + tr(L^{-1} + B^{-1}), \label{eq:3nodesTrP_1}
				\end{align}
				From these expressions, we see that the difference in the trace of the covariance matrix relies on $tr(A^{-1})$ and $tr(L^{-1})$.
				We can extend the expressions for $D$ and $L$ as:
				\begin{align*}
								A = \frac{N}{\lambda_2}\mathbb{E}[G_{1}' r_1 \times G_1^{'T} r_1] = \Gamma_A(q) N\frac{\sigma_1^2}{\lambda_2}, \\
								L = \frac{N}{\lambda_3}\mathbb{E}[G_2' r_2 \times G_2^{'T} r_2] = \Gamma_L (q) N\frac{\sigma_2^2}{\lambda_3},
				\end{align*}
				where $\Gamma_A (q)$ and $\Gamma_L (q)$ are filters associated with the covariance function of the vector signal $G_i' r_i$ for $i = 1, 2$. 
				Since $G_1(q) \equiv G_2(q)$ we have that $\Gamma_A \equiv \Gamma_L$. 
				In this way, if $\lambda_2 / \sigma_1^2 > \lambda_3 / \sigma_2^2$ then $tr(A^{-1}) > tr(L^{-1})$, implying that EMP II is more accurate.   
				Otherwise, EMP I is the more accurate. 
				Under Assumption \ref{assu:equally},  SNR21 is equal to SNR32, then we have that $tr(P_1) = tr(P_2)$. 
\end{pf}

The choice of EMP II over EMP I is equivalent to the decision of either exciting or measuring node $2$.  
As stated in this Theorem, this choice depends on whether SNR32 of EMP II is larger than SNR21 of EMP I  - see Definition \ref{def:SNRji}. 
If the user has control over the input energy, 
she can use (\ref{eq:snrEMP_1vsEMP_2}) as a tool to choose a value $\overline{\sigma}_2^2 > \sigma_1^2 \frac{\lambda_3}{\lambda_2}$, 
for which a better precision will be achieved by using EMP II. 
The same results in Theorem \ref{theo:EMPsIandIItr3nodes} were obtained in \cite{mapurunga2021sysid} for state space cascade networks. 

In conclusion, if the network is fully symmetrical then both EMPs give the same overall accuracy, 
which is due to the fact that the EMPs are ``mirrored’’ (See Definition \ref{def:mirror}) --
one is obtained from the other by changing each excitation for a measurement and vice-versa, 
except the sink and the source. 
Moreover, 
each module is identified more accurately when it is a direct module.
On the other hand, 
when the SNRs are not uniform, best overall accuracy is obtained by the EMP in which the direct module has a larger SNR. 
These are general principles that, as will be seen further ahead in this paper, also apply to more general networks.

\section{Four node cascade networks}
\label{sec:4nodes}

In this section we consider cascade networks composed of four nodes. 
From identifiability conditions presented in Corollary \ref{cor:IdentifiabilityConditions}, there are four different minimal EMPs from one to choose, listed below: 
\begin{enumerate}[I]
	\item $\mathcal{B} = \left\lbrace 1\right\rbrace; \mathcal{C} = \left\lbrace 2, 3, 4\right\rbrace$; \label{itemB1C234}
	\item $\mathcal{B} = \left\lbrace 1, 2, 3\right\rbrace; \mathcal{C} = \left\lbrace 4\right\rbrace$; \label{itemB123C4}
	\item $\mathcal{B} = \left\lbrace 1, 2\right\rbrace; \mathcal{C} = \left\lbrace 3, 4\right\rbrace$; \label{itemB12C34}
	\item $\mathcal{B} = \left\lbrace 1, 3 \right\rbrace; \mathcal{C} = \left\lbrace 2, 4 \right\rbrace$. \label{itemB13C24}
\end{enumerate} 

Minimal EMPs \ref{itemB1C234} and \ref{itemB123C4} in the four nodes case can be seen as an extension of the minimal EMPs I and II from the three nodes case, since they preserve a similar structure. 
For these two EMPs, we will show that the results from the previous section also apply for the four nodes case and more generally for any number of nodes.  

Now, adopting the same condition of previous section stated in Assumption \ref{assu:identicalG}, i.e. \emph{identical} transfer functions \(G_1 = G_2 = G_3 \triangleq G\).
Under this condition, the information matrix for the minimal EMP \ref{itemB1C234} becomes:

\begin{align}
	M_1 =  \left[\begin{matrix} 
		  A_{21} + B_{31} + C_{41} & B_{31} + C_{41} & C_{41} \\
			 B_{31} + C_{41} &  B_{31} + C_{41} & C_{41} \\
			 C_{41} & C_{41} & C_{41}
		\end{matrix}\right]			
		\label{eq:M14nodes}
,\end{align}
where
\begin{align}
				A_{ji} &:= \frac{1}{\lambda_{j}} \mathbb{E}[ G' r_i  \times G'^T r_i], \label{eq:Aji} \\
				B_{ji} &:= \frac{1}{\lambda_{j}} \mathbb{E}[ G' G r_i \times G'^T G r_i], \label{eq:Bji}\\
				C_{ji} &:= \frac{1}{\lambda_{j}} \mathbb{E}[ G' G G r_i \times  G'^T G G r_i]. \label{eq:Cji}
\end{align}

For minimal EMP \ref{itemB123C4} one can find in a similar way: 
\begin{align}
	M_2 = \left[\begin{matrix} 
			C_{41} & C_{41} & C_{41} \\
			C_{41} & C_{41} + B_{42} & C_{41} + B_{42} \\
			C_{41} & C_{41} + B_{42} & A_{43} + C_{41} + B_{42}
		\end{matrix}\right].
		\label{eq:M24nodes}
\end{align}

Information matrix of minimal EMP \ref{itemB12C34} is as follows:

\begin{align}
	M_3 = \left[\begin{matrix} 
			B_{31} + C_{41} & C_{41} + B_{31}                   & C_{41} \\
			C_{41} + B_{31} & C_{41} + B_{31} + B_{42} + A_{32} & C_{41} + B_{42} \\
			C_{41}          & C_{41} + B_{42}                   & C_{41} + B_{42}
		\end{matrix}\right].
		\label{eq:M34nodes}
\end{align}

Finally, minimal EMP \ref{itemB13C24} yields the information matrix:

\begin{align}
	M_4 = \left[\begin{matrix} 
			A_{21} + C_{41} & C_{41}  & C_{41} \\
			C_{41} & C_{41} & C_{41}  \\
			C_{41} & C_{41} &  C_{41} + A_{43}
		\end{matrix}\right].
		\label{eq:M44nodes}
\end{align}

If we additionally adopt the conditions stated in Assumption \ref{assu:equally}, 
we have that 
\(A_{ji} = A\), \(B_{ji} = B\) and \(C_{ji} = C\) in (\ref{eq:M14nodes})-(\ref{eq:M44nodes}).
By inverting 
the information matrix of each EMP we obtain the covariance matrices for each parameter estimates, which are displayed in Table \ref{tab:4nodecovpar}.

\begin{table}[h!]
				\centering
				\caption{Covariance of the parameter estimates for each EMP under Assumptions \ref{assu:identicalG} and \ref{assu:equally}.}
				\label{tab:4nodecovpar}
				\begin{adjustbox}{width=\columnwidth}
				\begin{tabular}{c | c c c}
								\toprule
								EMP/$\theta$ & $\hat{\theta}_1 $ & $\hat{\theta}_2$ & $\hat{\theta}_3$ \\
								\midrule
								I & $A^{-1}$ & $A^{-1} + B^{-1}$ & $B^{-1} + C^{-1}$\\
								II & $B^{-1} + C^{-1}$ & $A^{-1} + B^{-1}$ & $A^{-1}$\\
								III & $f(A, B, C)$ \footnotemark & $[A + [2B^{-1} + C^{-1}]^{-1}]^{-1}$ & $f(A, B, C)$  \\
								IV & $A^{-1}$ & $2A^{-1} + C^{-1}$ & $ A^{-1}$\\
								\bottomrule%
				\end{tabular}
				\end{adjustbox}
\end{table}
\footnotetext{$f(A, B, C) \triangleq [[A^{-1} + B^{-1}]^{-1} + [B^{-1} + C^{-1}]^{-1}   ]^{-1} $.}

Notice that EMPs I and II are symmetrical, that is, they yield the same results but in the reverse order.
For these two EMPs the key observations made in the previous section are also valid as can be seem from Table \ref{tab:4nodecovpar}.
The covariance matrix of $\hat{\theta}_1$ does not depend on additional measurements in EMP I, 
while $cov ( \hat{\theta}_3 )$ in EMP II does not depend on the first and second input signals. 
We will show later that this result also holds for an arbitrary number of nodes. 
If we consider the covariances of $\hat{\theta}_1$ and $\hat{\theta}_2$ from EMP I in the three nodes case, 
we can see from Table \ref{tab:4nodecovpar} that EMP I in the four nodes case does not improve the corresponding accuracy of $\hat{\theta}_1$ and $\hat{\theta}_2$ with respect to the three nodes counterpart. 
This implies that adding a new measurement (from three nodes to four nodes) for EMP I does not improve the precision of the parameter estimates under Assumptions \ref{assu:identicalG} and \ref{assu:equally}. 
The covariances of $\hat{\theta}_1$ and $\hat{\theta}_3$ are the same in EMP IV, which corresponds to $\hat{\theta}_1$ in EMP I and $\hat{\theta}_3$ in EMP II. 
This phenomenon is linked to how the location of excitations and measurements is distributed in the EMPs. 
We will show later in Section \ref{sec:Nnodes} when this effect happens for general cascade networks.  

Conversely to the three nodes cascade network case where all minimal EMPs lead to equivalent overall accuracy, 
in the four nodes case there is a minimal EMP that provides better precision. 
The next result shows that  minimal EMP III yields a smaller trace of the covariance matrix than EMPs I and II. 

\begin{Theo}
				Consider a 4-nodes cascade network with dynamic matrix (\ref{eq:GCasc}), for $ n = 4$. 
				Under Assumptions \ref{assu:identicalG} and \ref{assu:equally}, 
				minimal EMP \ref{itemB12C34} yields a smaller trace of covariance matrix than minimal EMPs \ref{itemB1C234} and \ref{itemB123C4}:  
				\begin{equation}
								tr(P_3) \le tr(P_1) = tr(P_2). \label{eq:trP_3leP_1}
				\end{equation}
				\label{theo:EMP3better}
\end{Theo}
\begin{pf}
				To prove (\ref{eq:trP_3leP_1}) we are going to compare the covariance of each module of both minimal EMPs in Table \ref{tab:4nodecovpar}.
				If the following conditions hold:
				\begin{align}
								A^{-1} &\succ (A + (2B^{-1} + C^{-1})^{-1})^{-1}, \label{eq:A<A+2B+C}\\ 
								A^{-1} + B^{-1} &\succ 
								[[A^{-1} + B^{-1}  ]^{-1} + [B^{-1} + C^{-1}]^{-1}   ]^{-1}, \label{eq:A+B>sum(ABBC)} \\ 
								B^{-1} + C^{-1} &\succ 
								[[A^{-1} + B^{-1}  ]^{-1} + [B^{-1} + C^{-1}]^{-1}   ]^{-1},   \label{eq:B+C>sum(ABBC)}
				\end{align}
				then $tr(P_1) >  tr(P_3)$.
								This follows from the implication: $A \succ  B \implies tr(A) >  tr(B)$, 
								with $A \succ  B$ in the sense that $A - B \succ  0$ is positive definite.
				The above conditions can be made equivalent to:
				\begin{align}
								A^{-1} (A^{-1} + B^{-1} + B^{-1} + C^{-1})^{-1} A^{-1} &\succ  0,  \label{eq:B<BX}\\ 
								A^{-1} + B^{-1} & \succ 0, \\ 
								B^{-1} + C^{-1} & \succ 0,
				\end{align}
				after some manipulation and using the fact that $A \succcurlyeq  B \iff A^{-1} \preccurlyeq B^{-1}$.  
				The last two conditions hold true since they define the covariance matrices of $\hat{\theta}_2$, $\hat{\theta}_3$ in Table \ref{tab:4nodecovpar} for EMP I. 
				Condition (\ref{eq:B<BX}) holds by definition and it was obtained using the matrix inversion lemma. 
\end{pf}

This result implies that under Assumptions \ref{assu:identicalG} and \ref{assu:equally} using EMP III is more advantageous than EMPs I and II.  
As for EMP IV there are values of the modules for which EMP IV can be made better than the others.  
EMP III will be more accurate than any  other minimal EMP if: $tr(P_1) \leq tr(P_4)$. 
From Table \ref{tab:4nodecovpar}, this could be achieved as:
\begin{align*}
				3 A^{-1} + C^{-1} &\succcurlyeq A^{-1} + 2 B^{-1} + C^{-1} \\
				A^{-1} &\succcurlyeq B^{-1} \iff B \succcurlyeq A 
.\end{align*}
From (\ref{eq:Aji})-(\ref{eq:Bji}) we notice that for ``large'' values of $G$ the above expression will hold true. 
The rationale is that $A$ is filtered by $G$ to produce $B$,
which implies that for ``large'' values of $G$ the expression $B - A$ will tend to be positive definite. 
In this scenario, EMP III will be the experimental setting that yields the smaller trace of the covariance matrix. 
The dependence of the estimates' accuracy  on the ``size'' of some modules come as no surprise as observed in \cite{mapurunga2021sysid}. 
To further investigate this property in a more general setting we will consider a numerical example using modules with FIR structure.
\begin{Ex}
				Consider a four nodes cascade network with FIR structured modules: 
				\begin{align}
								G_i(q) = \sum_{k = 0}^M g_{ik} q^{-k}.
				\label{eq:GjiFIR}
				\end{align}
				We have simulated 50,000 different networks and for each run we picked the minimal EMP with smallest trace of the covariance matrix.
				Each module was obtained as the impulse response of a discrete-time second order low pass Butterworth filter sampled at $1$ Hz and with normalized cutoff frequency randomly selected from $[0.1 , 0.4]$. 
				At each run the impulse response is truncated as in (\ref{eq:GjiFIR}) such that $|g_{il}| < 10^{-4}$ for $l > M$.  
				Five scenarios were analyzed: $\mathbb{S}_1$ where Assumptions \ref{assu:identicalG} and \ref{assu:equally} hold, the second $\mathbb{S}_2$ where only Assumption \ref{assu:equally} holds. 
				The remaining three scenarios are set under Assumption \ref{assu:equally}, but in each scenario the first parameter of one module is multiplied by ten. 
				For the third scenario $\mathbb{S}_3$	the selected module is the first one $G_1$, while we modified $G_2$ in the fourth $\mathbb{S}_4$, and $G_3$ in the fifth scenario $\mathbb{S}_5$.
				Table \ref{tab:ResFir1} displays how often each minimal EMP was selected as the best for all scenarios considered. 
				\begin{table}[h!]
								\centering
								\caption{How often minimal EMPs were selected for FIR cascade networks.}
								\label{tab:ResFir1}
								\begin{adjustbox}{width=\columnwidth}
								\begin{tabular}{ccccc}
											 	\toprule
											Scenario		& EMP I (\%)  & EMP II (\%)  & EMP III (\%) & EMP IV (\%) \\ \midrule
											$\mathbb{S}_1$ & 0 & 0  & 100 &   0 \\ 
											$\mathbb{S}_2$ & 0  & 0 & 96.65 & 3.35 \\ 
											$\mathbb{S}_3$ & 54.15 & 0  & 0 &   45.85 \\ 
											$\mathbb{S}_4$ & 0 & 0  & 100 &   0 \\ 
											$\mathbb{S}_5$ & 0 & 54.20  & 0 &   45.80 \\	\bottomrule 
								\end{tabular}
								\end{adjustbox}
				\end{table}
\end{Ex}

What this example shows is that, when all factors are equal as in scenario $\mathbb{S}_1$, EMP III provides the most accurate estimates. 
Even when we consider different modules, as in scenario $\mathbb{S}_2$, we have that EMP III still provides the best results in the vast majority of the cases. 
The remaining scenarios were designed to illustrate the ``size'' effect of some modules. 
In fact, in scenario $\mathbb{S}_3$ the only selected EMPs were I and IV, this happened because in these EMPs $G_1$ is a direct module - see Definition \ref{def:dirmod}. 
The same effect is behind the scenarios $\mathbb{S}_4$ -- where EMP III is the only one for which $G_2$ is a direct module -- and $\mathbb{S}_5$ for which $G_3$ is a direct module of EMPs II and IV.  
These results are expected since similar results were obtained for state-space cascade networks in \cite{mapurunga2021sysid}, which can be interpreted as first order FIR networks. 
In Section 6 we will explore more structures and we will observe similar results as those in Table \ref{tab:ResFir1}.

The dominance of EMP III with respect to the others is related to its structure.  
In this EMP, nodes near the source are excited, while nodes near the sink are measured. 
This structure represents a more balanced pattern since EMP III has equal shares of excitations and measurements and they are distributed 
such that half of the network provides information for the other half.
This kind of EMP will also have an advantage over others in a cascade network with $n$ number of nodes. 

Now if we relax Assumption  \ref{assu:equally} and consider Assumption \ref{assu:identicalG} only, one can get a similar result as Theorem \ref{theo:EMPsIandIItr3nodes} by inverting (\ref{eq:M14nodes}), (\ref{eq:M1_3nodes}), and (\ref{eq:M24nodes}). 

\begin{Theo}
				Consider a cascade network with four nodes and dynamic matrix as in (\ref{eq:GCasc}). 
			Under Assumption \ref{assu:identicalG} we have the following.
			\begin{enumerate}
							\item If SNR43 $>$ SNR21 and SNR42 $>$ SNR31,
											then EMP II yields more accurate results than EMP I. \label{item:EMPII<EMPI} 
							\item If SNR32 $>$ SNR43,
											then EMP III results in better accuracy than EMP I. \label{item:EMPIII<EMPI}
							\item If if SNR32 $>$ SNR 21, 
											then EMP III is more accurate than EMP II. \label{item:EMPIII<EMPII}
			\end{enumerate}	
\end{Theo}
\begin{pf}
				Let us start with item \ref{item:EMPII<EMPI}.  
				We first notice that under Assumption \ref{assu:identicalG} the following holds:
								$A_{ji} = \frac{\sigma_i^2}{\lambda_j} \Gamma_A,$ and 
								$B_{ji} = \frac{\sigma_i^2}{\lambda_j} \Gamma_B$. 	
								From (\ref{eq:M14nodes})-(\ref{eq:M44nodes}) and Table \ref{tab:4nodecovpar} we have that EMP II is more accurate than EMP I if:
			 \begin{align*}
							 A_{21} - A_{43} \succ 0 \iff \left(\frac{\sigma_1^2}{\lambda_2} - \frac{\sigma_3^2}{\lambda_4}\right) \Gamma_A \succ 0, \\ 
							 B_{31} - B_{42} \succ 0 \iff  \left(\frac{\sigma_1^2}{\lambda_3} - \frac{\sigma_2^2}{\lambda_4}\right) \Gamma_B \succ 0
			 ,\end{align*}	
			 from which follows the expressions for item \ref{item:EMPII<EMPI}. 
			 Now, items \ref{item:EMPIII<EMPI} and \ref{item:EMPIII<EMPII} can be proved in a similar fashion, for this reason we prove only the former. 
			 From the information matrix (\ref{eq:M34nodes}) and the expressions (\ref{eq:A<A+2B+C}) and (\ref{eq:A+B>sum(ABBC)}) in Theorem \ref{theo:EMP3better} we have that EMP III is more accurate than EMP I if:
			 \begin{align*}
							 A_{21}^{-1} \succ [A_{32} + [B_{31}^{-1} + B_{42}^{-1} + C_{41}^{-1}]^{-1}]^{-1} \iff \\
							 [B_{31}^{-1} + B_{42}^{-1} + C_{41}^{-1}]^{-1} + \left(\frac{ \sigma_2^2}{\lambda_3 } - \frac{\sigma_1^2}{\lambda_2}  \right)\Gamma_A \succ 0 \\ 
							 A_{21}^{-1} + B_{31}^{-1} \succ [[A_{32}^{-1} + B_{31}^{-1}]^{-1} + [B_{42}^{-1} + C_{41}^{-1}]^{-1}]^{-1} \iff \\
							 [[A_{32}^{-1} + B_{31}^{-1}]^{-1} - [A_{21}^{-1} + B_{31}^{-1}]^{-1} ] + [B_{42}^{-1} + C_{41}^{-1}]^{-1} \succ 0 \\
							 \implies 
							 \left( \frac{\sigma_2^2}{\lambda_3} - \frac{\sigma_1^2}{\lambda_2} \right) \Gamma_A  \succ 0 
			 .\end{align*}
			 From the above expression one can infer the conditions stated in this Theorem, since 
			 expression (\ref{eq:B+C>sum(ABBC)}) holds generally also in this case. 			 
\end{pf}

The first result is a direct extension of the result in Theorem \ref{theo:EMPsIandIItr3nodes}. 
These results reveal that the selection of the best EMP will also depend on the SNRji relationship among some nodes. 
It is worth noticing the comparison between EMPs III and I and also between EMPs III and II. 
In both cases, the key factor is the difference among the SNRji's of the direct modules of EMPs III ($G_2$), I ($G_1$), and II ($G_3$).

We have shown that the principles from the previous section are valid for the four node networks. 
As demonstrated in the last section, EMPs I and II provide equal overall accuracy, since they are mirrored versions of each other.  
However, in contrast with the case of networks with three nodes, EMP III yields better precision when compared to EMPs I and II. 
For FIR cascade networks EMP III dominates over all other minimal EMPs. 
A principle that emerges from this result is that EMP III yields better accuracy due to its uniform pattern with equal shares of excitations and measurements. 
In fact, this principle was already observed for state-space cascade networks in \cite{mapurunga2021sysid}, 
and it is also valid for cascade networks with more nodes as we will show in Section \ref{sec:Nnodes}. 
When the network is not uniformly excited, we have extended the result from the previous section and we have show that comparison among the EMPs I-II and III can be made based on the SNRji of the direct modules. 
The influence of the direct modules is clear in the case of FIR networks, when other factors are equal i. e. in a uniformly excited network, the EMPs with ``large'' direct modules have an advantage over the others.

\section{The general case}
\label{sec:Nnodes}

In the previous sections we considered cascade dynamic networks with just a few number of nodes.
Under Assumptions \ref{assu:identicalG} and \ref{assu:equally}, we have observed the following.
Minimal EMPs I and II  yield the same accuracy for cascade networks with three and four nodes. 
However, in the four nodes cases, EMP III outperforms EMPs I and II with respect to the trace of the covariance matrix.  
In this section, 
we will show that the phenomenon of yielding similar covariance matrices is not unique to EMPs I and II.  
We provide a result that characterizes all minimal EMPs with same overall accuracy. 
Furthermore, we will show that, as in the 4 nodes case, there is a minimal EMP that results in better estimates when compared to other minimal EMPs. 

The reason that minimal EMPs I and II yield, 
under Assumptions \ref{assu:identicalG} and \ref{assu:equally}, 
same overall accuracy is due to the symmetry of excitation and measurements. 
This happens because these EMPs are mirrored versions of each other.
For EMP I, we excite only the source and measure the other nodes, while for EMP II it holds the converse, only the sink is measured  and the other nodes are excited.
We now introduce the concept of mirrored EMP as follows.

\begin{Def}
				Consider an $n$-nodes cascade network, for which minimal EMP IX $\left(\mathcal{B}_{IX}, \mathcal{C}_{IX} \right) $ and minimal EMP XI $\left( \mathcal{B}_{XI}, \mathcal{C}_{XI} \right) $ apply. 
				Minimal EMP XI is a \emph{mirrored} version of minimal EMP IX if
				the set of excited and measured nodes are formed as
				$\mathcal{B}_{XI} = \{n-j+1 \,|\, j \in \mathcal{C}_{IX} \}  $ and 
				$\mathcal{C}_{XI} = \{ n-j+1 \,|\,j \in \mathcal{B}_{IX} \} $.
				\label{def:mirror}
\end{Def}

This definition implies that there is a symmetry with respect to the source and sink nodes for any minimal EMP and its mirrored version. 
If there is an excited (measured) node that is $k$ nodes ahead from the source, then in the \emph{mirrored} EMP the node that is $k$ nodes behind the sink must be measured (excited).   
Using this definition we see that EMP II is a mirrored version of EMP I and vice-versa. 
With this definition at hand we are in position to state the next result, which relates the accuracy of a given EMP and its mirrored version. 

\begin{Theo}
				Consider an $n$-nodes cascade network for which there are minimal EMPs that apply.
				Under Assumptions \ref{assu:identicalG} and \ref{assu:equally}, 
				a minimal EMP and its mirrored version have the same overall accuracy.  
\end{Theo}
\begin{pf}
				We are going to show that for a minimal EMP and its mirrored version, the information matrix can be written as:
				\begin{align}
								M = Q \overline{M} Q^T
								\label{eq:M=QMmQ}
				\end{align}
				where $\overline{M}$ is the information matrix associated with the mirrored version of an EMP and  
				\begin{align*}
								Q = \left[\begin{matrix}
										0 & \cdots & \cdots & 0 & 0 & I\\
										0 & \cdots & \cdots & 0 & I & 0\\
										0 & \cdots & \dots & I & 0 & 0\\
										I & \cdots & \cdots & 0 & 0 & 0\\
										\end{matrix}\right]
				.\end{align*}
				Notice that $Q = Q^T = Q^{-1}$ is a permutation matrix, thus both trace and determinant of $M$ and $\overline{M}$ are equal.
				The effect of pre and pos multiplying $Q$ is equivalent to reversing the order of the rows and columns of $M$. 
				Therefore, we just need to show that a mirrored version of an EMP has information matrix with reversed rows and columns. 
			  Now, we can decompose the gradient of the optimal predictor according to the transfer function from input $i$ to output $j$ as:
				\begin{align}
								\psi_{ji} = \left[ \begin{matrix} 
																\boldsymbol{0}_{i-1}\\
																\frac{ \partial \rho_{ji}}{\partial G_i} \\
																\vdots \\
																\frac{ \partial \rho_{ji}}{\partial G_{j - 1}} \\
																\boldsymbol{0}_{n - j }
								\end{matrix} \right] =
								\left[ \begin{matrix} 
																\boldsymbol{0}_{i-1}\\
																G_i' \frac{ \rho_{ji}}{ G_i} \\
																\vdots \\
																G'_{j - 1} \frac{\rho_{ji}}{ G_{j - 1}} \\
																\boldsymbol{0}_{n - j }
								\end{matrix} \right]
								\label{eq:psiji}
				\end{align}
				where $\rho_{ji} \triangleq \prod_{k = i}^{j-1} G_k r_i$, and $\boldsymbol{0}_k \in \mathbb{R}^{k  }$ is a vector filled with zeros. 
				Let $M_{ji} \triangleq \frac{1}{\lambda_j}\mathbb{E} \psi_{ji} \psi_{ji}^T$,
				the information matrix can be obtained as: $M = \sum_{j \in \mathcal{C}, i \in \mathcal{B}} M_{ji}$. 	
				From the structure of $\psi_{ji}$, the first $i-1$ block rows and columns of $M_{ji}$ are zero. 
				Similarly, the last $n-j$ block rows of $M_{ji}$ are also zero.
				Under Assumptions \ref{assu:identicalG} and \ref{assu:equally}, it holds that all nonzero elements of $\psi_{ji}$ are equal, which implies that the elements of $M_{ji}$ are the same. 
				Consider an arbitrary EMP defined by its set of excited nodes $\mathcal{B}$  and measured nodes $\mathcal{C}$. 
				For every pair $j \in {\mathcal{C}}$ and $i \in {\mathcal{B}}$, there is a reflected version of $\psi_{ji}$ in the mirrored EMP, such that:
				\begin{align*}
								\overline{\psi}_{n-i+1, n-j+1} = \left[ \begin{matrix} 
																\boldsymbol{0}_{n - j}\\
																G_{n-j+1}' \frac{ \rho_{n-i+1, n-j+1}}{ G_{n-j+1}} \\
																\vdots \\
																G_{n - i}' \frac{\rho_{n-i+1, n-j+1}}{ G_{j - 1}} \\
																\boldsymbol{0}_{i - 1}
								\end{matrix} \right] 
				.\end{align*}
				This means that $\overline{\psi}_{n-i+1, n-j+1}$ is a reversed, from top to bottom, version of $\psi_{ji}$.  
				From this relationship, we have that $\overline{M}_{n-i+1, n-j+1}$ can be written as $Q M_{ji} Q^T$. 
				Therefore, the information matrix of the mirrored EMP is as follows 
				\begin{align*}
								\overline{M} = \sum_{k \in \overline{\mathcal{C}}, l \in \overline{\mathcal{B}}} \overline{M}_{kl}= 
								\sum_{ j \in \mathcal{C}, i \in \mathcal{B}} Q M_{ji} Q^T = Q M Q^T 
				.\end{align*}
\end{pf}

This Theorem gives a framework for which one can exchange an excitation for a measurement (or the converse) without affecting the overall accuracy of the estimates. 
This property reveals a duality between excitation and measurement for which the key property is the symmetry of the EMPs for the cascade network. 
Among the EMPs there is always a mirrored EMP that yields the same accuracy under Assumptions \ref{assu:identicalG} and \ref{assu:equally}. 
However, for some minimal EMPs, the mirrored equivalent is not a different EMP but the EMP itself. 
This will be the case when the EMP is symmetrical with respect to the excitation and measurements. 
For cascade networks with an odd number of nodes, there are no minimal EMPs identical to their mirrored versions. 
Therefore, the number of minimal EMPs to be analyzed is halved, since a mirrored EMP produces the same accuracy.  
When the number of nodes in the network is even, there will be a total of $3 \cdot 2^{n - 4}$ minimal EMPs yielding different covariance matrices. 
The phenomenon of equal trace of covariance matrices for EMPs I and II in the case of three and four nodes is therefore a general result and valid for any $n$ nodes cascade network. 

In a similar way, the observations made in Section \ref{sec:3nodes} for minimal EMPs I and II can be further generalized. 
As we have seen in Section \ref{sec:3nodes} and \ref{sec:4nodes}, accuracy of the first module in EMP I is not improved by other measurements and its quality is related to the equality between the first two modules of the network. 
In \cite{Wahlberg2009CascAuto}, the authors have conjectured that a similar phenomenon would apply instead to the last module in EMP I, but they have shown that this does not happen.
However, an extension of this reasoning holds when we consider EMP II.
This conclusion is dependent on the EMP employed. 
The next result formalizes the above statements.  

\begin{Theo}
				Consider an n-nodes cascade network with dynamic matrix as in (\ref{eq:GCasc}),
				for which minimal EMPs apply. 
				The following holds:
				\begin{enumerate}
								\item The accuracy of $\hat{\theta}_1$ in any minimal EMP such that $2 \in \mathcal{C}$ and $G_1 \equiv G_2$ is not improved by any additional excitation signals or								 measurements; \label{item:theta1notimproved}  
								\item The accuracy of $\hat{\theta}_{n-1}$ in any minimal EMP such that $n-1 \in \mathcal{B}$ and $G_{n-2} \equiv G_{n-1}$ is not improved by any additional excitation signal or measurements. \label{item:thetannotimproved} 
				\end{enumerate}
\end{Theo}
\begin{pf}
				First, let us prove item \ref{item:theta1notimproved}.
				We decompose the gradient of the optimal predictor as in (\ref{eq:psiji}).
				Thus, the information matrix can be written as $
				M = \sum_{j \in \mathcal{C}, i \in \mathcal{B}} \frac{1}{\lambda_j} \mathbb{E} \psi_{ji} \psi_{ji}^T
				$. 
				Notice, however, that for any  $\psi_{ji}$ for $j, i \ge 3$ there is no dependence on $G_1$ and $G_2$. 
				The only terms that have the influence of $G_1$ and $G_2$ are $\psi_{k1}$ for $k = 2, \dots, n$.  
				Since we assume that $G_1 \equiv G_2$, we have that the first and second block rows of any $\psi_{ji}$ are exactly the same.   
				Therefore, the same holds for $M$ with exception of the first block element. 
				Since $2 \in \mathcal{C}$ the term $\psi_{21}$ appears in $M$ and is the only term which adds only to the first block element. 
				Thus, we can write $M$ as:
				\begin{align*}
								M = \left[\begin{matrix}
																A + X_1 & X_1 & \cdots & X_n \\
																X_1  & X_1 & \dots & X_n \\
																\vdots & \vdots & \tilde{M}_{11} & \tilde{M}_{12} \\
																X_n & X_n & \tilde{M}_{12}^T & \tilde{M}_{22}
								\end{matrix}\right],
				\end{align*}
				where $A \triangleq N / \lambda_2 \mathbb{E} G_1' r_1 G_1' r_1$. 
				Now, define:
				\begin{align*}
								Q \triangleq \left[\begin{matrix}
																I & -I & 0 \cdots & \cdots 0 \\
																0 & I & 0 \cdots & \cdots0 \\
																\vdots & \vdots & \tilde{Q}_{11} & \tilde{Q}_{12} \\
																0 & 0 & \tilde{Q}_{12}^T & \tilde{Q}_{22}
								\end{matrix}\right]
				.\end{align*}
				The covariance matrix can thus be obtained as $P_1 = Q^T \overline{M}^{-1} Q $, where
				\begin{align*}
								\overline{M}^{-1} = \left[\begin{matrix}
																A^{-1} & 0 \\
																0 & \tilde{S}_{22}
								\end{matrix}\right]
				.\end{align*}
				Thus, the accuracy of the first module is independent of the rest of the network signals under the conditions stated.   
				We can proceed similarly in the proof of item \ref{item:thetannotimproved}.	
				However, for this case, we have that the last two block rows are exactly the same, 
				since $G_{n-2} \equiv G_{n-1}$ and $n-2 \in \mathcal{B}$.
				The term $\psi_{n, n-1}$ adds only to the last block element in last block row.
				The corresponding covariance matrix can be obtained as:
				\begin{align*}
								P_2 = \left[\begin{matrix} 
																\tilde{P}_{11} & \tilde{P}_{12} \\
																\tilde{P}_{12}^T & A^{-1}
								\end{matrix}\right]
				,\end{align*}
				where $A \triangleq N / \lambda_{n} \mathbb{E} G'_{n-1} r_{n-1}  G^{'T}_{n-1} r_{n-1}$. 
				Therefore, the accuracy of the last module is independent of the network signals under the conditions stated.
\end{pf}

The results presented in this theorem come as no surprise. 
They can be interpreted as follows. 
The additional information about $G_1$ that the network signals could provide is only through the measurements. 
Thus, not even in the case when all signals are measured the accuracy of $G_1$ is improved when $G_1 \equiv G_2$ and $2 \in \mathcal{C}$. 
The dual situation happens with $G_{n-1}$ and the additional excitation signals.

We have shown in Theorem \ref{theo:EMP3better} that for a four nodes cascade network EMP III achieves better overall accuracy when compared to EMPs I and II. 
A principle that emerged from this result is that it would be better to excite the nodes near the source and measure the remaining nodes close to the sink. 
This observation will also be true for networks with an arbitrary number of nodes as shown in Section \ref{sec:Numerical}.  
Due to the complexity of the covariance expressions we will only provide an analytical result for 5-nodes networks. 
This result will naturally extend to cascade networks with more nodes.  

\begin{Theo}
				Consider a cascade network with five nodes. 
				Under Assumptions \ref{assu:identicalG} and \ref{assu:equally}, a smaller trace of the covariance matrix is obtained by EMP III: $(\{1, 2\}, \{3, 4, 5\} ) $ when compared to EMP I: $(\{1\}, \{2, 3, 4, 5\}  )$. 
\end{Theo}
\begin{pf}
				First we notice that the covariance of corresponding parameter estimates under EMP I is the same as shown in (\ref{eq:3nodesP_2})  and Table \ref{tab:4nodecovpar}. 
				More generally, one can show that the covariance of the parameter estimates for EMP I are: 
				\begin{align*}
								cov(\hat{\theta}^I_i) = X_i^{-1} + X_{i-1}^{-1}
				,\end{align*}
				where $X_i = \frac{1}{\lambda}\mathbb{E} G' \prod_{k = 0}^{i-1} G r_1 \times G'^T \prod_{l = 0}^{i-1} G r_1$. 
				Now, 
				we only need to show that:
				\begin{align*}
								cov (\hat{\theta}^{I}_1)  \succ	cov (\hat{\theta}^{III}_2),\, 
								cov (\hat{\theta}^{I}_2)  \succ	cov (\hat{\theta}^{III}_1),\\
								cov (\hat{\theta}^{I}_3)  \succ	cov (\hat{\theta}^{III}_3),\,
								cov (\hat{\theta}^{I}_4) \succ	cov (\hat{\theta}^{III}_4)
				.\end{align*}
				After lengthy calculations, this is equivalent to:
				\begin{align}
							&A^{-1} \succ \nonumber \\
											 &\;\;[A + [B^{-1} + [[B^{-1} + C^{-1}]^{-1} + [C^{-1} + D^{-1}]^{-1}]^{-1}]^{-1}]^{-1} 
											 \label{eq:A>A+B+B+C}\\
											 &	A^{-1} + B^{-1} \succ \nonumber\\
											 &[[A^{-1} + B^{-1}]^{-1} + [B^{-1} + C^{-1}]^{-1} + [C^{-1} + D^{-1}]^{-1}]^{-1}\\
											 & B^{-1} + C^{-1} \succ \nonumber\\
											 &[B + [Z_1^{-1} + [B + [C^{-1} + D^{-1}]^{-1}]^{-1}]^{-1}]^{-1}\\
											 & C^{-1} + D^{-1} \succ \nonumber \\
											 & [C + [Z_2^{-1} + [C + [A^{-1} + B^{-1}]^{-1}]^{-1}]^{-1} ]^{-1}
											 \label{eq:C+D>C+Z2}
				,\end{align}
				where $A$, $B$ and $C$ are defined as (\ref{eq:Aji})-(\ref{eq:Cji}), $D \triangleq \frac{N }{\lambda_5} \mathbb{E} G' G^{3} r \times G' G^3 r$ and  
				\begin{align*}
								0 \prec Z_1 &\triangleq C - [B + C][A + 2B + C]^{-1}[B + C],\\
								0 \prec Z_2 &\triangleq D - [C + D][B + 2C + D]^{-1}[C + D]
				.\end{align*}
				By applying the matrix inversion lemma on the right side of (\ref{eq:A>A+B+B+C})-(\ref{eq:C+D>C+Z2}) we get positive definite expressions by definition. 
\end{pf}

Now we have established that there is a minimal EMP which provides better accuracy than at least another minimal EMP. 
Since mirrored EMPs yield equal overall accuracy, the same conclusion is valid for the mirrored versions of EMPs stated in this Theorem. 
For a more general case where Assumptions \ref{assu:identicalG} and \ref{assu:equally} do not hold, we expect that these EMPs will tend to perform better. 
Indeed, this will be observed 
in the numerical results presented in the next section. 
We also point to the fact that EMP I applied to an $n$ nodes cascade network does not improve any estimates with respect to a $n-1$ nodes. 
This does not happen with other minimal EMPs, like those where the first nodes are excited and the last ones are measured.

In summary, we have established three principles that influence the accuracy of the modules' estimates in a cascade network. 
Firstly, we demonstrated that ``mirrored'' minimal EMPs provide the same overall accuracy. 
Therefore, we expect that when all quantities involved are arbitrary, there is no preferred choice between a particular EMP and its mirrored version. 
This choice will depend upon the magnitude of certain modules within the network and the SNRji of some nodes. 
Moreover, we demonstrated a topological principle for cascade networks: minimal EMPs where the nodes near the source are excited and the nodes close to the sink are measured yield most accurate results under Assumptions \ref{assu:identicalG} and \ref{assu:equally}. 
This principle was previously observed in \cite{mapurunga2021sysid} and it is once more confirmed for general cascade networks. 
Finally, the key observations of \cite{Wahlberg2009CascAuto} - see Section \ref{sec:3nodes}, were shown to be dependent not only on common dynamics between some modules but also on the experimental setting employed.

\section{Numerical Analysis}
\label{sec:Numerical}

In this section we analyze how the different factors presented so far work together and how they compare with each other. 
This is done through numerical experiments that demonstrate that the guiding principles developed until now also apply to the general case of $n$-nodes networks.

For the numerical experiments we consider that the following will be valid in all experiments.
A total of 10,000 network simulations will be performed. 
In each run we consider cascade networks with cardinality from four to eight nodes.
All signals involved are realizations of Gaussian white noise processes. 
				The input $\{ r_i(t)\}$ is zero-mean Gaussian with variance $\sigma_i^2$ for $i \in \mathcal{B}$, 
				while the corrupting noise $\{ e_j(t)\}$ is also zero mean process, but with variance $\lambda_j$ for $j \in \mathcal{C}$. 
A new realization of the random signals involved is performed at each run of the simulation.
With respect to the EMPs, we will consider two scenarios:
\begin{enumerate}[(i)]
				\item Assumption \ref{assu:equally} holds. 
				For this scenario we have chosen $\sigma_i^2 = 1,\; \forall i \in \mathcal{B}$ and $\lambda_j = 0.01,\; \forall j \in \mathcal{C}$.
				\label{item:scenarioequally}	
\item For the second analyzed scenario the variances $\sigma_i^2$ and $\lambda_j$ will be drawn from a uniform distribution $\mathcal{U}\left(0.001, 50\right)$ . 
				\label{item:scenariorandom}
\end{enumerate}

We remark that the numerical values of the SNR in the first scenario do not influence the decision of the best minimal EMP, 
since in this case the choice depends only on the numerical parameters and the EMP itself. 
For each cardinality of the network we will choose at every run a best EMP, the one with smallest trace of the covariance matrix. 
Due to space restrictions, we are going to list just the two best minimal EMPs for each cardinality. 
With respect to the structure of each model we will consider two structures: first and second order transfer functions.  
The first order module is parametrized as: 
\begin{align}
				G_i(q, \theta_i) = \frac{b_i}{q + a_i}
				\label{eq:FOTF}
,\end{align}
where $\theta_i = [a_i\; b_i]^T$. 
In each run of the numerical simulation each parameter is randomly selected. 
Each module's parameter  $a_i$ is sampled from $\mathcal{U}\left(0.1, 0.9\right)$, while 
$b_i$'s are sampled from $\mathcal{U}(0.5, 2) $. 
The results obtained for the first order transfer function (\ref{eq:FOTF}) under scenarios (i) and (ii) are displayed in Table \ref{tab:ResFOTF1},  
which shows the frequency in which the two most selected EMPs were chosen as the best for cascade networks with number of nodes from  4 up to 8. 
\begin{table}[h!]
				\centering
				\caption{How often the two best minimal EMPs$(\mathcal{B}, \mathcal{C})$ were selected considering first order modules for scenario (i) - under Assumption \ref{assu:equally} - and (ii) where all quantities were randomly selected. }
				\label{tab:ResFOTF1}
				\begin{adjustbox}{width=\columnwidth}
				\begin{tabular}{cccccc}
								 \toprule
								 $n$ & Scen. &best EMP & \% & runner-up EMP & \% \\ \midrule
								 \multirow{2}{*}{4}  & (i) &$(\{1, 2\},  \{3, 4\}  )$ & 54.12 & $(\{1, 3\}, \{2, 4\}) $ & 21.25 \\ 
																		 & (ii)  &$(\{1, 2\},  \{3, 4\}  )$ & 50.22 & $(\{1, 3\}, \{2, 4\}) $ & 18.55 \\ \midrule

								 \multirow{2}{*}{5}  & (i) &$(\{1, 2\}, \{3, 4, 5\})$ & 24.79 &  $(\{1, 2, 3\}, \{4, 5\})$   & 24.72 \\
																		 & (ii) &$(\{1, 2, 3\}, \{4, 5\})$ & 23.58 &  $(\{1, 2\}, \{2, 4, 5\})$   & 23.1 \\ \midrule

								 \multirow{2}{*}{6}  & (i) &$(\{1, 2, 3\}, \{4, 5, 6\})$ & 18.88 & $(\{1, 2, 4\}, \{3, 5, 6\})$ & 16.72 \\
																		 & (ii) &$(\{1, 2, 3\}, \{4, 5, 6\})$ & 17.08 & $(\{1, 2, 4\}, \{3, 5, 6\})$ & 13.55 \\ \midrule

								 \multirow{2}{*}{7}  & (i) &$(\{1, 2, 3, 4\}, \{5, 6, 7\})$ & 12.47 & $(\{1, 2, 3\}, \{ 4, 5, 6, 7\})$ & 11.78 \\
																		 & (ii) &$(\{1, 2, 3, 4\}, \{5, 6, 7\})$ & 10.56 & $(\{1, 2, 3\}, \{ 4, 5, 6, 7\})$ & 10.36 \\ \midrule

								 \multirow{2}{*}{8}  & (i) &$(\{1, 2, 3, 4 \}, \{5, 6, 7, 8\})$ & 10.81  & $(\{1, 2, 3\}, \{4, 5, 6, 7, 8\})$ & 8.3  \\ 
																		 & (ii) &$(\{1, 2, 3, 4 \}, \{5, 6, 7, 8\})$ & 9.15  & $(\{1, 2, 3\}, \{4, 5, 6, 7, 8\})$ & 7.64  \\ \bottomrule

				\end{tabular}
\end{adjustbox}
\end{table}

The results from this Table show that for all network cadinalities the best minimal EMP was the one where the nodes near the source were excited and the nodes near the sink were measured.
Thus, the results obtained for cascade networks with four nodes are also valid for larger networks even when Assumptions \ref{assu:identicalG} and \ref{assu:equally} do not hold.

For small cardinalities, such as four nodes, the difference in frequency was more than two times with respect to the runner-up EMP, while for bigger cardinalities a slightly increase in the frequency was observed. 
Notice that the best EMP and the runner up for five and seven nodes are mirrored EMPs and they provide similar accuracy. 
This means that together they account for almost half of the selections for fives nodes and a fifth for seven nodes case. 
The decrease in percentage of the best EMP when we increase the cardinality of the network is also due to the large number of available minimal EMPs, 
for instance, for a network with cardinality eight there are a total of 64 minimal EMPs to choose.     

Once we have observed that exciting the first nodes and measuring the last ones is the best approach, we wonder by how much the best EMP yields better precision than the others. 
This is a crucial aspect since one could benefit simply from choosing the structure of excitations and measurements in the network. 
To answer this question we have compared the ratio of the trace of covariance matrices obtained by the best EMP and the runner-up and also the ratio between the best EMP and the worst EMP for the results displayed in Table {\ref{tab:ResFOTF1}}.
We depict in Table \ref{tab:ResFOTFmedian} the median of the ratio best/runner up and best/worst for the two scenarios analyzed.  

\begin{table}[h!]
				\centering
				\caption{Median of the trace of covariance matrix ratio between the best EMP and the runner up and between best EMP and worst EMP for first order modules.}
				\label{tab:ResFOTFmedian}
				\begin{tabular}{c | c c | c c }
				\toprule 
								&  \multicolumn{2}{c}{(i) Under Assumption 2.2 } & \multicolumn{2}{c}{(ii) All random} \\ \midrule
								n & runner-up & worst EMP & runner-up & worst EMP\\
								4 & 1.59 & 10.23  & 1.96 & 15.45    \\
								5 & 1.31 & 26.14  & 1.54 & 46.40   \\
								6 & 1.19 & 56.64 & 1.36 & 111.69  \\
								7 & 1.12 & 111.42 & 1.24 & 240.86  \\
								8 & 1.08 & 227.18 & 1.19 & 555.80   \\ \bottomrule
				\end{tabular}
\end{table}

We see in this Table that for small network cardinalities (four and five nodes) we can have from  30\% up to almost double precision improvement compared to the runner-up minimal EMP.
Whereas the situation for larger cardinalities we have improvements of at least 8\% (for the eight nodes case). 
The situation dramatically changes when the best EMP is compared to the EMP that yielded worst accuracy. 
For a network with few nodes we have at least ten times better precision and for larger networks this number grows even bigger to 555 (for eight nodes). 
The large difference observed for cardinalities with more than six nodes is partly due to well-known fact that when we increase the number of parameters the variance also increases. 
Therefore, for larger networks it is even more important to not choose an EMP arbitrarily. 

Now, for the second order transfer function we adopt the following structure:
\begin{align}
				G_i(q, \theta_i) = \frac{\theta_{i 1} q + \theta_{i 2}}{q^2 + \theta_{i 3} q + \theta_{i 4}}
.\end{align}
The poles of $G_i(q, \theta_i)$  are randomly selected from the right side of the unitary disk, 
while the zeros are drawn from a disk with radius three. 
We have performed the same experiments that we have done for first order modules. 
Similar results to Table \ref{tab:ResFOTF1} are presented in Table \ref{tab:ResSOTF}. 

\begin{table}[h!]
				\centering
				\caption{How often the best minimal EMPs$(\mathcal{B}, \mathcal{C})$  were selected for second order modules under scenarios (i) (Assumption \ref{assu:equally}) and (ii).}
				\label{tab:ResSOTF}
				\begin{adjustbox}{width=\columnwidth}
					\begin{tabular}{cccccc}
								 \toprule
								 $n$ & Scen. &best EMP & \% & runner-up EMP & \% \\ \midrule
								 \multirow{2}{*}{4}  & (i) &$(\{1, 2\},  \{3, 4\}  )$ & 57.38 & $(\{1, 3\}, \{2, 4\}) $ & 17.7 \\ 
																		 & (ii) &$(\{1, 2\},  \{3, 4\}  )$ & 49.7 & $(\{1\}, \{2, 3, 4\}) $ & 18.09 \\ \midrule

								 \multirow{2}{*}{5}  & (i) &$(\{1, 2, 3\}, \{4, 5\})$ & 29.06 &  $(\{1, 2\}, \{3, 4, 5\})$   & 28.64 \\
																		 & (ii) &$(\{1, 2\}, \{3, 4, 5\})$ & 27.55 &  $(\{1, 2, 3\}, \{4, 5\})$   & 23.86 \\ \midrule

								 \multirow{2}{*}{6}  & (i) &$(\{1, 2, 3\}, \{4, 5, 6\})$ & 24.36 & $(\{1, 2, 4\}, \{3, 5, 6\})$ & 17.76 \\
																		 & (ii) &$(\{1, 2, 3\}, \{4, 5, 6\})$ & 18.78 & $(\{1, 2, 4\}, \{3, 5, 6\})$ & 12.83 \\ \midrule

								 \multirow{2}{*}{7}  & (i) &$(\{1, 2, 3\}, \{4, 5, 6, 7\})$ & 13.56 & $(\{1, 2, 3, 4\}, \{5, 6, 7\})$ & 13.17 \\
																		 & (ii) &$(\{1, 2, 3\}, \{4, 5, 6, 7\})$ & 11.05 & $(\{1, 2, 3, 4\}, \{ 5, 6, 7\})$ & 10.2 \\ \midrule

								 \multirow{2}{*}{8}  & (i) &$(\{1, 2, 3, 4 \}, \{5, 6, 7, 8\})$ & 10.6  & $(\{1, 2, 3, 5\}, \{4, 6, 7, 8\})$ & 6.85  \\ 
																		 & (ii) &$(\{1, 2, 3, 4 \}, \{5, 6, 7, 8\})$ & 7.22  & $(\{1, 2, 3\}, \{4, 5, 6, 7, 8\})$ & 5.33  \\ \bottomrule

				\end{tabular}

\end{adjustbox}
\end{table}

As can be observed from this Table, the results observed for first order modules are also valid for second order modules. 
Once more, there is a minimal EMP that is selected more often than others, 
the one where nodes near the source are excited and the nodes near the sink are measured. 
These results suggest that the principles derived from the analysis under Assumptions \ref{assu:identicalG} and \ref{assu:equally} can be applied as guidelines for the selection of the best EMPs.

We also have analyzed the gains in accuracy of the best EMP compared to the runner up EMP and the worst EMP, which are displayed in Table \ref{tab:ResSOTFprec}. 

\begin{table}[h!]
				\centering
				\caption{Median of the trace of covariance matrix ratio between the best EMP and the runner up and between the best EMP and worst EMP for second order modules. }
				\label{tab:ResSOTFprec}
				\begin{tabular}{c | c c | c c }
				\toprule 
								&  \multicolumn{2}{c}{(i) Under Assumption 2.2} & \multicolumn{2}{c}{(ii) All random} \\ \midrule
								n & runner-up & worst EMP & runner-up & worst EMP\\
								4   & 1.72 & 7.46   & 2.09 & 11.72  \\
								5   & 1.52 & 17.57  & 1.75 & 31.35 \\
								6   & 1.44 & 37.10  & 1.61 & 71.29 \\
								7   & 1.39 & 72.41  & 1.51 & 153.80  \\
								8   & 1.37 & 136.27 & 1.47 & 288.92   \\ \bottomrule
				\end{tabular}
\end{table}

Once again, we have observed that  the selection of the minimal EMP is crucial in the precision of the parameter estimates. 
In the case of second order modules the difference from the runner up is even larger than in the first order module, ranging from at least 37\% better to more than two times the precision. 

From the thousands of numerical experiments performed we have seen that the principles derived for the analytical results are also observed in the more general case where the network is arbitrarily excited and with different modules.  
Firstly, mirrored EMPs tend to have a similar performance on the accuracy of the estimates. 
Secondly we have observed that minimal EMPs where the first half of the nodes are excited while the remaining nodes are measured tend to give most accurate results. 
For any minimal EMP there are competing factors that will influence the decision of which EMP provides the most accurate estimates. 
On one hand, there is the influence of the SNR at some nodes on the precision of the parameter estimates. 
On the other hand, the magnitude of the modules' parameters may be a key decider for accuracy of the estimates. 
In any case, the structure of the excitations and measurements plays a major role in the selection of the best EMP as evidenced by the numeric examples.  

\section{Conclusion}
\label{sec:Conclusion}

In this work we have investigated how the allocation of excitations and measurements influences the accuracy of the parameter estimates obtained by the prediction error method in a linear cascade network.
A variance analysis 
was carried out to determine which EMP yields the most accurate estimates 
using the minimum number of excitations and measurements combined.
Accuracy was assessed through the trace of the Cramer-Rao lower bound matrix of the parameters' estimates. 

We have established
a number of key factors that influence the accuracy in cascade networks based either on  analytical results or on an extensive numerical analysis, or both. 
These factors together form fundamental principles that an experiment design should account for when the objective is to decide which EMP yields the most accurate results. 
The first factor is a topological principle that states that EMPs where the first half nodes are excited and the remaining nodes are measured yield the most accurate estimates. 
Secondly, a large signal-to-noise ratio should be applied in the direct modules of the EMPs. 
If, in addition to that, some prior knowledge is available, then the user should choose EMPs for which direct modules have a large magnitude. 
Thirdly, we have shown that some EMPs result in the same overall accuracy, which
allows the user to exchange excitations for measurements and vice-versa without losing precision of the estimates.  

Last, but not least, a very important finding is the large difference observed in the precision of the estimates when the best excitation and measurement pattern is compared to other candidates. This attests to the paramount importance of the topic in the experiment design phase, and serves as motivation for future research aiming at extending the results in this paper to more general network
topologies and other identification scenarios. 

\bibliographystyle{model5-names}
\biboptions{authoryear} 
\bibliography{VACDynet}

\begin{thebibliography}{32}
\expandafter\ifx\csname natexlab\endcsname\relax\def\natexlab#1{#1}\fi
\providecommand{\url}[1]{\texttt{#1}}
\providecommand{\href}[2]{#2}
\providecommand{\path}[1]{#1}
\providecommand{\DOIprefix}{doi:}
\providecommand{\ArXivprefix}{arXiv:}
\providecommand{\URLprefix}{URL: }
\providecommand{\Pubmedprefix}{pmid:}
\providecommand{\doi}[1]{\href{http://dx.doi.org/#1}{\path{#1}}}
\providecommand{\Pubmed}[1]{\href{pmid:#1}{\path{#1}}}
\providecommand{\bibinfo}[2]{#2}
\ifx\xfnm\relax \def\xfnm[#1]{\unskip,\space#1}\fi
\bibitem[{Bazanella et~al.(2017)Bazanella, Gevers, Hendrickx \&
  Parraga}]{Bazanella2017WhichNodes}
\bibinfo{author}{Bazanella, A.}, \bibinfo{author}{Gevers, M.},
  \bibinfo{author}{Hendrickx, J.}, \& \bibinfo{author}{Parraga, A.}
  (\bibinfo{year}{2017}).
\newblock \bibinfo{title}{{Identifiability of dynamical networks: Which nodes
  need be measured?}}
\newblock In {\it \bibinfo{booktitle}{IEEE Conference on Decision and
  Control}\/} (pp. \bibinfo{pages}{5870--5875}).
\bibitem[{Bazanella et~al.(2019)Bazanella, Gevers \&
  Hendricks}]{bazanella_network_2019}
\bibinfo{author}{Bazanella, A.~S.}, \bibinfo{author}{Gevers, M.}, \&
  \bibinfo{author}{Hendricks} (\bibinfo{year}{2019}).
\newblock \bibinfo{title}{Network identiﬁcation with partial excitation and
  measurement}.
\newblock In {\it \bibinfo{booktitle}{{IEEE} {Conference} on {Decision} and
  {Control}}\/} (p.~\bibinfo{pages}{7}).
\newblock \bibinfo{address}{Nice, France}: \bibinfo{publisher}{IEEE}.
\bibitem[{Cheng et~al.(2021)Cheng, Shi \& Van~den Hof}]{cheng_allocation_2021}
\bibinfo{author}{Cheng, X.}, \bibinfo{author}{Shi, S.}, \&
  \bibinfo{author}{Van~den Hof, P.~M.} (\bibinfo{year}{2021}).
\newblock \bibinfo{title}{Allocation of {Excitation} {Signals} for {Generic}
  {Identifiability} of {Linear} {Dynamic} {Networks}}.
\newblock {\it \bibinfo{journal}{IEEE Transactions on Automatic Control}\/},
  (pp. \bibinfo{pages}{1--1}).
\bibitem[{Dankers et~al.(2015)Dankers, Van~den Hof, Bombois \&
  Heuberger}]{dankers_errors--variables_2015}
\bibinfo{author}{Dankers, A.}, \bibinfo{author}{Van~den Hof, P. M.~J.},
  \bibinfo{author}{Bombois, X.}, \& \bibinfo{author}{Heuberger, P. S.~C.}
  (\bibinfo{year}{2015}).
\newblock \bibinfo{title}{Errors-in-variables identification in dynamic
  networks — {Consistency} results for an instrumental variable approach}.
\newblock {\it \bibinfo{journal}{Automatica}\/},  {\it \bibinfo{volume}{62}\/},
  \bibinfo{pages}{39--50}.
\bibitem[{Dankers et~al.(2016)Dankers, {Van den Hof}, Bombois \&
  Heuberger}]{Dankers2016PredInputSel}
\bibinfo{author}{Dankers, A.~G.}, \bibinfo{author}{{Van den Hof}, P. M.~J.},
  \bibinfo{author}{Bombois, X.}, \& \bibinfo{author}{Heuberger, P. S.~C.}
  (\bibinfo{year}{2016}).
\newblock \bibinfo{title}{{Identification of Dynamic Models in Complex Networks
  With Prediction Error Methods: Predictor Input Selection}}.
\newblock {\it \bibinfo{journal}{IEEE Transactions on Automatic Control}\/},
  {\it \bibinfo{volume}{61}\/}, \bibinfo{pages}{937--952}.
\bibitem[{Dimovska \& Materassi(2021)}]{dimovska_control_2021}
\bibinfo{author}{Dimovska, M.}, \& \bibinfo{author}{Materassi, D.}
  (\bibinfo{year}{2021}).
\newblock \bibinfo{title}{A {Control} {Theoretic} {Look} at {Granger}
  {Causality}: {Extending} {Topology} {Reconstruction} to {Networks} {With}
  {Direct} {Feedthroughs}}.
\newblock {\it \bibinfo{journal}{IEEE Transactions on Automatic Control}\/},
  {\it \bibinfo{volume}{66}\/}, \bibinfo{pages}{699--713}.
\bibitem[{Gevers \& Bazanella(2015)}]{Gevers2015ExpdsgIssues}
\bibinfo{author}{Gevers, M.}, \& \bibinfo{author}{Bazanella, A.}
  (\bibinfo{year}{2015}).
\newblock \bibinfo{title}{{Identification in dynamic networks: identifiability
  and experiment design issues}}.
\newblock In {\it \bibinfo{booktitle}{IEEE Conference on Decision and
  Control}\/} \bibinfo{number}{Cdc} (pp. \bibinfo{pages}{4005--4010}).
\bibitem[{Gevers et~al.(2017)Gevers, Bazanella \&
  Parraga}]{Gevers2017IdentifiDyNet}
\bibinfo{author}{Gevers, M.}, \bibinfo{author}{Bazanella, A.}, \&
  \bibinfo{author}{Parraga, A.} (\bibinfo{year}{2017}).
\newblock \bibinfo{title}{{On the identifiability of dynamical networks}}.
\newblock {\it \bibinfo{journal}{IFAC-PapersOnLine}\/},  {\it
  \bibinfo{volume}{50}\/}, \bibinfo{pages}{10580--10585}.
\bibitem[{Gevers et~al.(2019)Gevers, Bazanella \&
  Pimentel}]{gevers_identifiability_2019}
\bibinfo{author}{Gevers, M.}, \bibinfo{author}{Bazanella, A.~S.}, \&
  \bibinfo{author}{Pimentel, G.~A.} (\bibinfo{year}{2019}).
\newblock \bibinfo{title}{Identifiability of {Dynamical} {Networks} {With}
  {Singular} {Noise} {Spectra}}.
\newblock {\it \bibinfo{journal}{IEEE Transactions on Automatic Control}\/},
  {\it \bibinfo{volume}{64}\/}, \bibinfo{pages}{2473--2479}.
\bibitem[{Gevers et~al.(2018)Gevers, Bazanella \&
  da~Silva}]{gevers_practical_2018}
\bibinfo{author}{Gevers, M.}, \bibinfo{author}{Bazanella, A.~S.}, \&
  \bibinfo{author}{da~Silva, G.~V.} (\bibinfo{year}{2018}).
\newblock \bibinfo{title}{A practical method for the consistent identification
  of a module in a dynamical network}.
\newblock {\it \bibinfo{journal}{IFAC-PapersOnLine}\/},  {\it
  \bibinfo{volume}{51}\/}, \bibinfo{pages}{862--867}.
\bibitem[{Gracy et~al.(2021)Gracy, Paré, Sandberg \&
  Johansson}]{gracy_analysis_2021}
\bibinfo{author}{Gracy, S.}, \bibinfo{author}{Paré, P.~E.},
  \bibinfo{author}{Sandberg, H.}, \& \bibinfo{author}{Johansson, K.~H.}
  (\bibinfo{year}{2021}).
\newblock \bibinfo{title}{Analysis and distributed control of periodic epidemic
  processes}.
\newblock {\it \bibinfo{journal}{IEEE Transactions on Control of Network
  Systems}\/},  {\it \bibinfo{volume}{8}\/}, \bibinfo{pages}{123--134}.
\bibitem[{Han et~al.(2020)Han, Wik, Kersten, Dong \&
  Zou}]{Han_nextgeneration_2020}
\bibinfo{author}{Han, W.}, \bibinfo{author}{Wik, T.}, \bibinfo{author}{Kersten,
  A.}, \bibinfo{author}{Dong, G.}, \& \bibinfo{author}{Zou, C.}
  (\bibinfo{year}{2020}).
\newblock \bibinfo{title}{Next-generation battery management systems: Dynamic
  reconfiguration}.
\newblock {\it \bibinfo{journal}{IEEE Industrial Electronics Magazine}\/},
  {\it \bibinfo{volume}{14}\/}, \bibinfo{pages}{20--31}.
\bibitem[{Hendrickx et~al.(2019)Hendrickx, Gevers \&
  Bazanella}]{Hendrickx2018PartialNodes}
\bibinfo{author}{Hendrickx, J.~M.}, \bibinfo{author}{Gevers, M.}, \&
  \bibinfo{author}{Bazanella, A.~S.} (\bibinfo{year}{2019}).
\newblock \bibinfo{title}{{Identifiability of Dynamical Networks With Partial
  Node Measurements}}.
\newblock {\it \bibinfo{journal}{IEEE Transactions on Automatic Control}\/},
  {\it \bibinfo{volume}{64}\/}, \bibinfo{pages}{2240--2253}.
\bibitem[{Jin et~al.(2021)Jin, Yuan \& Gonçalves}]{jin_full_2021}
\bibinfo{author}{Jin, J.}, \bibinfo{author}{Yuan, Y.}, \&
  \bibinfo{author}{Gonçalves, J.} (\bibinfo{year}{2021}).
\newblock \bibinfo{title}{A {Full} {Bayesian} {Approach} to {Sparse} {Network}
  {Inference} {Using} {Heterogeneous} {Datasets}}.
\newblock {\it \bibinfo{journal}{IEEE Transactions on Automatic Control}\/},
  {\it \bibinfo{volume}{66}\/}, \bibinfo{pages}{3282--3288}.
\newblock \bibinfo{note}{Conference Name: IEEE Transactions on Automatic
  Control}.
\bibitem[{Legat \& Hendrickx(2020)}]{legat_local_2020}
\bibinfo{author}{Legat, A.}, \& \bibinfo{author}{Hendrickx, J.~M.}
  (\bibinfo{year}{2020}).
\newblock \bibinfo{title}{Local {Network} {Identifiability} with {Partial}
  {Excitation} and {Measurement}}.
\newblock In {\it \bibinfo{booktitle}{2020 59th {IEEE} {Conference} on
  {Decision} and {Control} ({CDC})}\/} (pp. \bibinfo{pages}{4342--4347}).
\bibitem[{Ljung(1999)}]{ljung1998system}
\bibinfo{author}{Ljung, L.} (\bibinfo{year}{1999}).
\newblock {\it \bibinfo{title}{System Identification: Theory for the User}\/}.
\newblock \bibinfo{publisher}{Pearson Education}.
\bibitem[{Mapurunga \&
  Bazanella(2021{\natexlab{a}})}]{mapurunga_identifiability_2021}
\bibinfo{author}{Mapurunga, E.}, \& \bibinfo{author}{Bazanella, A.~S.}
  (\bibinfo{year}{2021}{\natexlab{a}}).
\newblock \bibinfo{title}{Identifiability of {Dynamic} {Networks} from
  {Structure}}.
\newblock {\it \bibinfo{journal}{IFAC-PapersOnLine}\/},  {\it
  \bibinfo{volume}{54}\/}, \bibinfo{pages}{55--60}.
\bibitem[{Mapurunga \& Bazanella(2021{\natexlab{b}})}]{mapurunga2021sysid}
\bibinfo{author}{Mapurunga, E.}, \& \bibinfo{author}{Bazanella, A.~S.}
  (\bibinfo{year}{2021}{\natexlab{b}}).
\newblock \bibinfo{title}{Optimal {Allocation} of {Excitation} and
  {Measurement} for {Identification} of {Dynamic} {Networks}}.
\newblock {\it \bibinfo{journal}{IFAC-PapersOnLine}\/},  {\it
  \bibinfo{volume}{54}\/}, \bibinfo{pages}{43--48}.
\bibitem[{Pukelsheim(2006)}]{pukelsheimOptimalDesignExperiments2006}
\bibinfo{author}{Pukelsheim, F.} (\bibinfo{year}{2006}).
\newblock {\it \bibinfo{title}{Optimal design of experiments}\/}.
\newblock \bibinfo{publisher}{SIAM}.
\bibitem[{Ramaswamy et~al.(2021)Ramaswamy, Bottegal \& Van~den
  Hof}]{ramaswamy_learning_2021}
\bibinfo{author}{Ramaswamy, K.~R.}, \bibinfo{author}{Bottegal, G.}, \&
  \bibinfo{author}{Van~den Hof, P. M.~J.} (\bibinfo{year}{2021}).
\newblock \bibinfo{title}{Learning linear modules in a dynamic network using
  regularized kernel-based methods}.
\newblock {\it \bibinfo{journal}{Automatica}\/},  {\it
  \bibinfo{volume}{129}\/}, \bibinfo{pages}{109591}.
\bibitem[{Ramaswamy \& Van~den Hof(2020)}]{ramaswamy_local_2020}
\bibinfo{author}{Ramaswamy, K.~R.}, \& \bibinfo{author}{Van~den Hof, P.~M.}
  (\bibinfo{year}{2020}).
\newblock \bibinfo{title}{A local direct method for module identification in
  dynamic networks with correlated noise}.
\newblock {\it \bibinfo{journal}{IEEE Transactions on Automatic Control}\/},
  (pp. \bibinfo{pages}{1--1}).
\bibitem[{{Van den Hof} et~al.(2013){Van den Hof}, Dankers, Heuberger \&
  Bombois}]{VanDenHof2013ComplexPem}
\bibinfo{author}{{Van den Hof}, P. M.~J.}, \bibinfo{author}{Dankers, A.~G.},
  \bibinfo{author}{Heuberger, P.~S.}, \& \bibinfo{author}{Bombois, X.}
  (\bibinfo{year}{2013}).
\newblock \bibinfo{title}{{Identification of dynamic models in complex networks
  with prediction error methods - Basic methods for consistent module
  estimates}}.
\newblock {\it \bibinfo{journal}{Automatica}\/},  {\it \bibinfo{volume}{49}\/},
  \bibinfo{pages}{2994--3006}.
\bibitem[{Van~Waarde et~al.(2019)Van~Waarde, Tesi \&
  Camlibel}]{van_waarde_necessary_2019}
\bibinfo{author}{Van~Waarde, H.}, \bibinfo{author}{Tesi, P.}, \&
  \bibinfo{author}{Camlibel, M.~K.} (\bibinfo{year}{2019}).
\newblock \bibinfo{title}{Necessary and {Sufficient} {Topological} {Conditions}
  for {Identifiability} of {Dynamical} {Networks}}.
\newblock {\it \bibinfo{journal}{IEEE Transactions on Automatic Control}\/},
  (pp. \bibinfo{pages}{4525 -- 4537}).
\bibitem[{Van~Waarde et~al.(2021)Van~Waarde, Tesi \&
  Camlibel}]{van_waarde_topology_2021}
\bibinfo{author}{Van~Waarde, H.~J.}, \bibinfo{author}{Tesi, P.}, \&
  \bibinfo{author}{Camlibel, M.~K.} (\bibinfo{year}{2021}).
\newblock \bibinfo{title}{Topology identification of heterogeneous networks:
  {Identifiability} and reconstruction}.
\newblock {\it \bibinfo{journal}{Automatica}\/},  {\it
  \bibinfo{volume}{123}\/}, \bibinfo{pages}{109331}.
\bibitem[{Wahlberg et~al.(2009)Wahlberg, Hjalmarsson \&
  M{\aa}rtensson}]{Wahlberg2009CascAuto}
\bibinfo{author}{Wahlberg, B.}, \bibinfo{author}{Hjalmarsson, H.}, \&
  \bibinfo{author}{M{\aa}rtensson, J.} (\bibinfo{year}{2009}).
\newblock \bibinfo{title}{{Variance results for identification of cascade
  systems}}.
\newblock {\it \bibinfo{journal}{Automatica}\/},  {\it \bibinfo{volume}{45}\/},
  \bibinfo{pages}{1443--1448}.
\bibitem[{Wahlberg et~al.(2007)Wahlberg, Jansson, Matsko \&
  Molander}]{Wahlberg2007}
\bibinfo{author}{Wahlberg, B.}, \bibinfo{author}{Jansson, M.},
  \bibinfo{author}{Matsko, T.}, \& \bibinfo{author}{Molander, M.~A.}
  (\bibinfo{year}{2007}).
\newblock \bibinfo{title}{Experiences from subspace system identification -
  comments from process industry users and researchers}.
\newblock (pp. \bibinfo{pages}{315--327}).
\newblock \bibinfo{publisher}{Springer}.
\bibitem[{Weerts et~al.(2018{\natexlab{a}})Weerts, Galrinho, Bottegal,
  Hjalmarsson \& Van~den Hof}]{weerts_sequential_2018}
\bibinfo{author}{Weerts, H.}, \bibinfo{author}{Galrinho, M.},
  \bibinfo{author}{Bottegal, G.}, \bibinfo{author}{Hjalmarsson, H.}, \&
  \bibinfo{author}{Van~den Hof, P.} (\bibinfo{year}{2018}{\natexlab{a}}).
\newblock \bibinfo{title}{A sequential least squares algorithm for {ARMAX}
  dynamic network identification}.
\newblock {\it \bibinfo{journal}{IFAC-PapersOnLine}\/},  {\it
  \bibinfo{volume}{51}\/}, \bibinfo{pages}{844--849}.
\bibitem[{Weerts et~al.(2018{\natexlab{b}})Weerts, Van~den Hof \&
  Dankers}]{weerts_identifiability_2018}
\bibinfo{author}{Weerts, H.}, \bibinfo{author}{Van~den Hof, P. M.~J.}, \&
  \bibinfo{author}{Dankers, A.} (\bibinfo{year}{2018}{\natexlab{b}}).
\newblock \bibinfo{title}{Identifiability of linear dynamic networks}.
\newblock {\it \bibinfo{journal}{Automatica}\/},  {\it \bibinfo{volume}{89}\/},
  \bibinfo{pages}{247--258}.
\bibitem[{Weerts et~al.(2018{\natexlab{c}})Weerts, Van~den Hof \&
  Dankers}]{weerts_prediction_2018}
\bibinfo{author}{Weerts, H.~H.}, \bibinfo{author}{Van~den Hof, P.~M.}, \&
  \bibinfo{author}{Dankers, A.~G.} (\bibinfo{year}{2018}{\natexlab{c}}).
\newblock \bibinfo{title}{Prediction error identification of linear dynamic
  networks with rank-reduced noise}.
\newblock {\it \bibinfo{journal}{Automatica}\/},  {\it \bibinfo{volume}{98}\/},
  \bibinfo{pages}{256--268}.
\bibitem[{Weerts et~al.(2015)Weerts, Dankers \& Van~den
  Hof}]{weerts_identifiability_2015}
\bibinfo{author}{Weerts, H. H.~M.}, \bibinfo{author}{Dankers, A.~G.}, \&
  \bibinfo{author}{Van~den Hof, P. M.~J.} (\bibinfo{year}{2015}).
\newblock \bibinfo{title}{Identifiability in dynamic network identification}.
\newblock {\it \bibinfo{journal}{IFAC-PapersOnLine}\/},  {\it
  \bibinfo{volume}{48}\/}, \bibinfo{pages}{1409--1414}.
\bibitem[{Weerts et~al.(2020)Weerts, Linder, Enqvist \& Van~den
  Hof}]{weerts_abstractions_2020}
\bibinfo{author}{Weerts, H. H.~M.}, \bibinfo{author}{Linder, J.},
  \bibinfo{author}{Enqvist, M.}, \& \bibinfo{author}{Van~den Hof, P. M.~J.}
  (\bibinfo{year}{2020}).
\newblock \bibinfo{title}{Abstractions of linear dynamic networks for input
  selection in local module identification}.
\newblock {\it \bibinfo{journal}{Automatica}\/},  {\it
  \bibinfo{volume}{117}\/}, \bibinfo{pages}{108975}.
\bibitem[{Ye et~al.(2011)Ye, Guy-Bart, Sean \& Jorge}]{YUAN20111230}
\bibinfo{author}{Ye, Y.}, \bibinfo{author}{Guy-Bart, S.},
  \bibinfo{author}{Sean, W.}, \& \bibinfo{author}{Jorge, G.}
  (\bibinfo{year}{2011}).
\newblock \bibinfo{title}{Robust dynamical network structure reconstruction}.
\newblock {\it \bibinfo{journal}{Automatica}\/},  {\it \bibinfo{volume}{47}\/},
  \bibinfo{pages}{1230--1235}.

\end{thebibliography}

\end{document}